\documentclass[floatfix,epsf,
prb,twocolumn,
amsmath,amssymb,showpacs,superscriptaddress]{revtex4}
\usepackage[english]{babel}
\usepackage{graphics}
\usepackage{amsmath}
\usepackage{dsfont}
\usepackage{graphicx}
\usepackage[figtopcap]{subfigure}
\usepackage{enumerate}
\begin{document}\setlength{\unitlength}{1mm}
\newcommand{\ud}{\mathrm{d}}
\newcommand{\kvec}[2]{\begin{pmatrix} #1 \\ #2 \end{pmatrix} }
\newcommand{\rvec}[2]{\begin{pmatrix}#1 & #2 \end{pmatrix} }
\newcommand{\matt}[4]{\begin{pmatrix} #1 & #2 \\ #3 & #4 \end{pmatrix} }
\newcommand{\pdje}[1]{\partial_{#1}}
\newcommand{\ve}{\varepsilon}
\title{Edge magnetoplasmons in a wide armchair graphene ribbon with a weak superlattice potential:
 finite frequency gaps and zero group velocity}

\author{O. G. Balev}
\address{Departmento de Fisica, Universidade Federal do Amazonas, 69077-000, Manaus, Brazil\\}
\author{A. C. A. Ramos}
\address{Universidade Federal do Cear\'{a}, Campus Cariri, 63040-360, Juazeiro do Norte, Cear\'{a}, Brazil\\}

\author{H. O. Frota}
\address{Departmento de Fisica, Universidade Federal do Amazonas, 69077-000, Manaus, Brazil\\}
\begin{abstract}	
We show strong effects of a weak and smooth, on the magnetic length, superlattice potential upon 
edge magnetoplasmons (EMPs) at the armchair edge,  
with a smooth steplike electrostatic lateral confining potential,
of a wide graphene channel in  the $\nu=2$ quantum Hall effect  regime.
The superlattice potential leads to essential enlargement of a number of 
EMPs, descend from two fundamental EMPs in the absence of superlattice. 
For the wave vector $k_{x}$ within the first Brillouin zone, the EMPs show
as the regions of acoustical or quasi-acoustical dispersion, with a finite value of group velocity,
so the regions with frequency gaps, where a group velocity is nullified at some $k_{x}$.
We obtain that for $k_{x} \to 0$ only for two EMPs the frequency tends to zero
as for other EMPs it obtains finite values.
Strong dependence of dispersion relations of the EMPs from the period of the superlattice $a_{0}$ and the distance $d$ 
from a metallic gate is shown; in particular, for typical size of a gap,  for characteristic value of the 
frequency and $k_{x}$ at which the group velocity is reduced to zero.
At the frequency that corresponds to zero group velocity of pertinent fundamental EMP branch the response of 
the system should present a strong resonance.  
 \end{abstract}
\maketitle

\section{Introduction}\label{sec1}
Graphene after experimental discovery of its high-quality freestanding samples, \cite{novo}
 has attracted a strong attention.\cite{cast}
 Charge carriers in a single-layer graphene possess  a gapless,  linear spectrum close to the 
$K$ and $K'$ points \cite{novo,cast,wallace} and manifest behavior of  chiral massless particles  with a
"light speed" equal to the Fermi velocity, $v_{F}$.
Graphene shows a lot of unusual effects, e.g.: the Klein paradox\cite{cast,klein,kat}, i.e., the perfect 
transmission through arbitrarily high and wide barriers upon normal
incidence (as far as a Dirac-type Hamiltonian is valid), a half-integer quantum Hall effect (QHE) \cite{cast,brey,aba,gus0}, and a zitterbewegung, \cite{cast,zit1,zit2}
i.e., effect induced by a lateral confinement of Dirac fermions. Properties of the latter effect are essentially 
modified by a strong magnetic field. Extra Dirac points in the energy spectrum for superlattices in graphene
have been obtained if the amplitude of periodic potential is sufficiently large while its period is small enough.
\cite{park2009,brey2009,barbier2010} In particular, this leads to new properties of the QHE.\cite{park2009}

Graphene's edges have also been studied considerably, \cite{cast,brey,aba,gus0,gus,milt}
in particular, in connection with the QHE \cite{cast,brey,aba,gus0}; for some phenomena it matters 
a type, the armchair or zigzag, of edges. \cite{cast,brey,aba,gus0}
Edge magnetoplasmons (EMPs) in graphene have been studied only recently; \cite{balev2011}
it is shown that in the $\nu=2$ QHE regime at the armchair edge, and in the presence of a smooth steplike electrostatic
lateral confining potential, the chirality, spectrum, spatial structure, and number of the fundamental EMPs depend strongly
on the position of the Fermi level $E_{F}$.

In the case (i) of Ref. \onlinecite{balev2011}, when $E_{F}$ intesects  (see Fig. 1, cf. with Fig. 1(a) 
of Ref. \onlinecite{balev2011}) four degenerate states of the zero LL 
at one location and two degenerate states of this LL at a different location,
two fundamental EMPs are present: counterpropagating and with essential spatial overlap. 
This is in contrast with EMPs in conventional two-dimensional electron systems (2DES) which 
give only one fundamental EMP at the $\nu=2$ QHE regime, with negligible spin-splitting;
for conventional 2DES different types of EMPs have been studied theoretically\cite{volkov91,aleiner94,wen91,stone92,bal,bal2000,bal99} 
and experimentally. \cite{ashoori92,ernst96,kukushkin09}
Above two counterpropagating EMPs can be on resonance if a strong coupling of the EMPs holds at the ends of the segment 
$L_{x}^{em} \leq L_{x}$, where $L_{x}$ is the length of graphene channel.\cite{balev2011} 

In present study for the case  qualitatively outlined in Fig. 1  (i.e., it is the case (i) of Ref. \onlinecite{balev2011})
 we explore theoretically effect of a weak and smooth superlattice potential   
$V_{s}(x)=V_{s}\cos(Gx)$ with  $G=2\pi/a_{0}$, upon EMPs. Here, in agreement with speculations of \cite{balev2011} 
that a strong Bragg coupling is possible due to a weak superlattice along 
the edge (with period $L_{x}^{em}$, if $L_{x}/L_{x}^{em} \gg 1$), we show that
$V_{s}(x)$ can have a strong effect on two fundamental EMPs leading to manifestation of 
resonance effects; in particular, referred to in the abstract.
Present EMPs in graphene with the superlattice are very different from the EMPs treated previously for
conventional 2DES with a superlattice. \cite{bal2000b}

In Sec. II A we obtain the wave functions and the spectra of LLs in an infinitely
large graphene flake in the presence of a perpendicular magnetic field  and  of a smooth
electrostatic confining potential, along the $y$ direction, as without 
$V_{s}(x)$ so in its presence.
In Sec. II B we study the combined effect of a  smooth, step-like electrostatic confining potential
and of  armchair graphene edges, at $y=\pm L_{y}/2$,  and of the superlattice potential
$V_{s}(x)$ on the local Hall conductivity
in the $\nu=2$ QHE regime.  In Sec. III we obtain strong renormalization of the EMPs in graphene by a weak superlattice
potential.  We make concluding remarks   in Sec. IV.

\begin{figure}[ht]
\vspace*{-0.5cm}
\includegraphics [height=10cm, width=8cm]{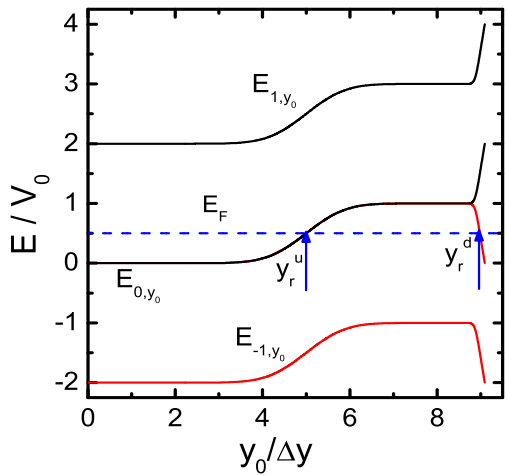}
\vspace*{-3.0cm}
\caption{(Color online) Energy spectrum of the $n=0, \pm 1$ LLs
as a function of the quantum number $y_{0}$, at the right half of symmetric graphene
channel with  armchair edges and the
smooth electrostatic potential, Eq. (\ref{eq17}),
for the Fermi level $E_{F}=V_{0}/2$. 
Spatially separated edge states
are created at  $y_{r}^{u}$ and at $y_{r}^{d}$ (marked by an upward arrow)
as the branches of the $n=0$ LL cross $E_{F}$.
The $\nu=2$ QHE is manifested in dc magnetotransport.
}
\end{figure}

\section{Graphene Channel and local Hall conductivity}\label{sec2}
\subsection{Effect of a smooth potential and of a weak periodic potential on the LLs}

We  consider a long and a wide flat graphene flake of length $L_{x}$ and width $L_{y}$, with armchair edges at $y=\pm L_{y}/2$
in the presence of a perpendicular magnetic field  ${\bf B}=B\hat{z}$,  of a smooth confining potential $V_y=V(y)$ along the
$y$ direction of electrostatic origin, and of a one-dimensional (1D) periodic potential $V_{s}(x)$ along the $x$ direction.
We assume that $V_{s}(x)=V_{s}\cos(Gx)$ is a weak 1D modulation potential of period $a_{0}$.
For definiteness we assume that the potential $V_y$ is symmetric.

If it is not otherwise stated, we  consider solutions with energy and wave vector close to the K point;
we present pertinent results for the energies and wave vectors
close to the K$^\prime$ point (valley) as well.  In the nearest-neighbor, tight-binding model the  one-electron Dirac Hamiltonian, for massless electrons, is $\mathcal{H} = \mathcal{H}_{0}+ \mathds{1} V_{s}(x)$ where
$\mathcal{H}_{0} = v_F \vec{\sigma}\cdot \hat{\vec{p}} + \mathds{1} V_y$, with $\mathds{1}$ the $2\times 2$ unit matrix.
Explicitly $\mathcal{H}_{0} $ is given by ($e > 0$)
\begin{equation}
    \mathcal{H}_{0} = v_F\matt{V_y/v_F}{p_x-i p_y-eBy}{p_x+i p_y-eBy}{V_y/v_F},
\label{eq1}
\end{equation}
where $p_x$ and  $p_y$ are components of the  momentum operator ${\bf p}$  and $v_F \approx 10^6 m/s$ the Fermi velocity. The vector potential is taken in the Landau gauge, ${\bf A}=(-By,0,0)$.

\subsubsection{Landau levels for a smooth potential $V(y)$}

First, we present properties of the LLs in the absence of periodic potential, when only
a smooth potential and armchair termination are assumed \cite{balev2011}.
The equation $(\mathcal{H}_{0} - E)\psi = 0$ admits solutions of the form
\begin{equation}
    \psi^{(0)}({\bf r})= e^{i k_{x \alpha} x}\Phi(y)/\sqrt{L_x},\quad \Phi(y) = \kvec{A\Phi_A(y)}{B\Phi_B(y)} ,
\label{eq2}
\end{equation}
where the components $\Phi_A(y)$ and $\Phi_B(y)$ correspond to the two sublattices and the coefficients A and B satisfy the relation $|A|^2+|B|^2=1$; ${\bf r}=\{x,y\}$.
Introducing the magnetic length $\ell_0=(\hbar/eB)^{1/2}$, $y_0=\ell_{0}^2k_{x \alpha}$,  the variable $\xi=(y-y_0)/\ell_0$,
and assuming that $V_y$ is a smooth function of $y$, with a characteristic scale $\Delta y \gg \ell_{0}$,
it follows \cite{balev2011} at $r=\ell_0 a/\hbar v_F \ll 1$ for the $n=0$ LL that
the energy $E_{0,k_{x \alpha}}^{(0)}=V(y_{0})$ and
 \begin{eqnarray}
&&\hspace*{-0.9cm}\Phi_{A\kappa }^{0}(\xi)=\Phi_{B\kappa }^{0}(\xi)=1/(\pi \ell_{0}^{2} )^{1/4}
e^{-\xi^2/2}, \nonumber \\*
&&A_\kappa^{0}=(1/r)[1-\kappa(1-r^2)^{1/2}]B_\kappa^{0} ,
\label{eq3}
\end{eqnarray}
where $a=\partial V/ \partial \xi|_{\xi=0}$, and $\kappa=+(-)$ corresponds to the K (K$^\prime$) valley.
here $B_{+}^{0}=A_{-}^{0}\approx 1$ and $A_{+}^{0}=B_{-}^{0}\approx r/2 \ll 1$.

Finally, without periodic potential for  any $n=0, \pm 1,\pm 2,...$ LL  and $y_{0}$ not too close to the
graphene lattice termination at $y=\pm L_{y}/2$ (see Fig. 1), the
eigenvalues $E_{n, k_{x \alpha}}^{(0),\kappa}=E_{n, y_0}^{(0),\kappa}$ can be written as
\begin{equation}
E_{n, k_{x \alpha}}^{(0),\kappa}=sgn(n) \frac{\hbar v_F}{\ell_0}\sqrt{2|n|} +V(y_0),\quad n=0,\pm 1,...
\label{eq4}
\end{equation}
where the sign function $sgn(n)=1$ and $-1$ for $n>0$ and $n<0$, respectively.
Notice that each $n \neq 0$ LL is twice degenerate with respect to the valley quantum number $\kappa$.
However, the latter index is kept in the left hand side of Eq. (\ref{eq4})  as for $y_{0}(k_{x \alpha})$ close to
the armchair edge these eigenvalues, and especially strongly $E_{0, k_{x \alpha}}^{(0),\kappa}$, become dependent on $\kappa$;
notice, for these conditions $\kappa$ can not be related only to one valley \cite{brey,aba,gus0}. 
Accordingly, for $n \neq 0$ LL  and $y_{0}$ not too close to the graphene lattice termination, see Fig. 1, the
eigenvalues (\ref{eq4}) are four times degenerate.  Wave functions pertinent to the eigenvalues Eq. (\ref{eq4})
we will denote also as $\psi^{(0),\kappa}_{n,k_{x \alpha}}({\bf r})$.

\subsubsection{Effect of potential $V(y)$ and of a weak smooth periodic potential $V_{s}(x)$ on the LLs}

Assuming $\hbar v_{F}/\ell_{0} \gg \hbar v_{g}(k_{x \alpha}) G \gg V_{s}/2$ and $G\ell_{0} \ll 1$ where the group velocity
$v_{g}(k_{x \alpha})=\hbar^{-1} dE^{(0)}_{0,k_{x \alpha}}/d k_{x \alpha}$, now we will study
how a weak periodic potential $V_{s}(x)$ modifies energies and
wave functions of the LLs for a smooth potential $V(y)$; in particular, for conditions pertinent to Fig. 1.
We calculate the eigenvalues and eigenfunctions corresponding to the
Hamiltonian $\mathcal{H} = \mathcal{H}_{0}+ \mathds{1} V_{s}(x)$ using the perturbation theory \cite{landaulif}.
Because further there will be important only the eigenstates localized, along $y$, near a right edge of the channel,
we will write formulas assuming, in particular, such eigenstates.
Similar with Ref. \cite{bal2000b}, we can neglect by a small ``nonresonance'' contributions, $n_{\beta} \neq n_{\alpha}$.
Then keeping only the ``resonance'' contributions, $n_{\beta}=n_{\alpha}$, and the terms of the first order over
$V_{s}$, e.g., for the eigenfunctions of the $n=0$ LL we obtain
\begin{equation}
\psi^{\kappa}_{0,k_{x \alpha}}({\bf r})=\psi^{(0),\kappa}_{0,k_{x \alpha}}
+\frac{V_{s}}{2\hbar v_{g}(k_{x \alpha}) \; G}
\Big[\psi^{(0),\kappa}_{0,k_{x \alpha}-G}
-\psi^{(0),\kappa}_{0,k_{x \alpha}+G}\Big] .
\label{eq5}
\end{equation}
Further, the eigenvalues are well approximated by the zero-order terms, i.e., $E_{n, k_{x \alpha}}^{\kappa} \approx E_{n, k_{x \alpha}}^{(0),\kappa}$,
where, for $y_{0}$ not too close to the graphene lattice termination, $E_{n, k_{x \alpha}}^{(0),\kappa}$ are given by Eq. (\ref{eq4}).
Indeed, the first order corrections are exactly nullified and the second order ones
are very small, e.g.: $\sim (V_{s}/2V_{0})^{2} \ll 1$ for the model potential Eq. (\ref{eq17}).

\subsection{Local Hall conductivity in the $\protect\nu=2$ QHE regime}

Extending magnetotransport formulas for the local Hall conductivity $\sigma_{yx}({\bf r})$ of a standard  2DES in the channel,
in the presence of a smooth lateral potential \cite{bee,thouless93,bal96},   we obtain, for linear responses
and in strong magnetic fields, $\sigma_{yx}({\bf r})$ in the form \cite{zhe}
\begin{equation}
\sigma_{yx}({\bf r})= n({\bf r}) e/B ,
\label{eq14}
\end{equation}
where the local electron density $n({\bf r})$ is smooth, on the characteristic scale $\ell_{0}$, as along
$y$, mainly monotonic, so along $x$, with a weak periodic modulation. It is given by
\begin{equation}
n({\bf r})=  \sum_{\alpha\kappa}f_{\alpha\kappa}\langle\alpha\kappa| \mathds{1}\delta({\bf r}-\hat{{\bf r}})|\alpha\kappa\rangle ,
\label{eq15}
\end{equation}
with $\alpha=\{n,k_{x \alpha}\}$;  $\sigma_{yx}(y)=-\sigma_{xy}(y)$. So far in Eqs. (\ref{eq14})-(\ref{eq15})
only the electrons from the conduction band LLs are assumed. However, if the valence band LLs
can essentially contribute to $\sigma_{yx}({\bf r})$, as for Fig. 1, the local hole density $p({\bf r})$
will contribute to the right hand part of Eq. (\ref{eq14}) by changing $n({\bf r})$ on $[n({\bf r})-p({\bf r})]$.
Then, for conditions relevant to Fig. 1, when only $(n=0, \kappa=\pm)$ LLs can essentially contribute to $\sigma_{yx}({\bf r})$
or a diagonal component of the local conductivity tensor,
equation (\ref{eq14}) is rewritten as
\begin{eqnarray}
\nonumber
\sigma_{yx}(x,y)&=&\frac{2e^{2}L_{x}}{h}
\sum_{\kappa=\pm} \int_{-\infty}^{\infty} dy_{0} [f_{0,y_{0},\kappa}- \delta_{\kappa,-}] \\
&&\times \left\langle \psi^{\kappa}_{0,k_{x \alpha}}({\bf r})\left\|\right.\psi^{\kappa}_{0,k_{x \alpha}}({\bf r})\right\rangle,
\label{eq16}
\end{eqnarray}
where $f_{n,y_{0},\kappa}$ is the Fermi function, the two-component column spinor wave function
$\left|\psi^{\kappa}_{0,k_{x \alpha}}({\bf r})\right\rangle$ is given by Eq. (\ref{eq5});
the factor $2$ accounts for spin degeneracy.
In Eqs. (\ref{eq15}),(\ref{eq16})  $\kappa=\pm$ is understood as the pseudospin
quantum number \cite{balev2011}; e.g., for $y_{0}>0$ only at $(L_y/2-y_0)/\ell_0 \gg 1$ it can be well approximated as the valley index.
A strong splitting between the electron, $\kappa=+$, and the hole, $\kappa=-$, branches
of the $n=0$ LL  \cite{cast,milt,brey}, due to
hybridization of the valley states
take place nearby the armchair edge, at $|L_y/2-y_0| \leq \ell_0$.
The eigenvalues of the $n=0$ LL for
$\kappa=+(-)$ increase (decrease) with increasing $y_{0}$.
However, for the $n \geq 1$ LLs the $\kappa=\pm$ branches at the armchair edge
have a small splitting, due to hybridization of the valley states, as their eigenvalues increase with increasing $y_{0}$;
these branches are attributed to the electron band.
Notice, the electron, $(n=0, \kappa=+)$, LL stems from the conduction band and the hole, $(n=0, \kappa=-)$, LL
arises from the valence band.

We now consider the situations depicted
in Fig. 1 for a wide symmetric armchair graphene ribbon  $L_{y}>2y_r \gg \Delta y \gg \ell_0$. For definiteness 
the smooth lateral potential is assumed as follows
\begin{equation}
\hspace*{-0.2cm}V(y)=
(V_0/2)\Big[2+\Phi ((y-y_r)/\Delta y) +\Phi ((y+y_r)/\Delta y)\Big],
\label{eq17}
\end{equation}
where $\Phi (x)$ 
is the probability integral. In Fig. 1 we have
$L_{y}=18 \Delta y$, $V_{0}=\hbar v_{F}/\sqrt{2} \ell_0$, $y_{r}=5 \Delta y$, and
$\Delta y=10\ell_{0}$. When the Fermi level $E_F$ is between the bottoms of the $n=0$ and $n=1$ LLs,
at $y_{0}=0$, and the condition $V_0\gg 2k_BT$ holds, the occupation of the $n \geq 1$ LLs is negligible; the same holds for
the $n=0$ LL in the regions of $y_{0}$ that are well above $E_F$,
 see Fig. 1. In addition to the smoothness of the potential Eq.(\ref{eq17}), we assume armchair edges 
of the graphene sheet at $y=\pm L_y/2$,
which cause the bending of the  LLs, \cite{cast,milt,brey}
and $L_y/2-y_r\geq \Delta y$. 
For conditions of Fig. 1 and qualitatively similar, the dc magnetotransport measurements will manifest the $\nu=2$ QHE.

Further, for conditions qualitatively similar with those of Fig. 1, in agreement with speculations of Ref. \cite{balev2011} we will show that
a weak periodic potential $V_{s}(x)$ can have a strong effect on two fundamental EMPs leading to manifestation of the
resonance effects. First, for convenience of a reader, we will present expressions for the local Hall conductivity,
obtained in Ref. \cite{balev2011}, that are pertinent to $V_{s}(x) \to 0$.

\subsubsection{Effect of a smooth potential, an armchair edge and a weak periodic potential
on local Hall conductivity in the $\protect\nu=2$ QHE regime}

 Now, for different regions of the graphene channel we will present expressions as for, obtained in Ref. \cite{balev2011},
the unperturbed
local Hall conductivity, $\sigma_{yx}^{(0)}(y)$,
 so for the main contribution induced by a finite $V_{s}(x)$, $\sigma_{yx}^{(1)}(x,y) \propto V_{s}$.
The superscript in $\sigma_{yx}^{(1)}(x,y)$  indicates that it is of the first order over $V_{s}$.
Correspondingly, this contribution in Eq. (\ref{eq16}) stems
from the first order contributions to the wave function Eq. (\ref{eq5}); 
to calculate $\sigma_{yx}^{(1)}(x,y)$ we will need also assume that $V_{s}/k_{B}T \ll 1$.
Point out that $\sigma_{yx}^{(0)}(y)$ is given by the right hand side of Eq. (\ref{eq16}) if to substitute
$\left|\psi^{\kappa}_{0,k_{x \alpha}}({\bf r})\right\rangle$, calculated by taking into account the first order
corrections, by the zero order wave function $\left|\psi^{(0),\kappa}_{0,k_{x \alpha}}({\bf r})\right\rangle$.

In   case (i), for $y_{0}>0$ and $(y_{r}^{d}-y_{0})/\ell_{0} \gg 1$, from
Eqs. (\ref{eq3})-(\ref{eq4}), (\ref{eq16})-(\ref{eq17}) it follows \cite{balev2011} 
\begin{equation}
\sigma_{yx}^{(0)}(y)=\frac{2e^2}{h} \tanh\left(\frac{V(y_r^u)-V(y)}{2k_{B} T}\right) ,
\label{eq18}
\end{equation}
where $V(y)$ is so smooth on the  scale of $\ell_{0}$ that  $\ell_{0} dV(y_{r}^{u})/dy \ll k_{B} T$;
the factor $4$ accounts  for spin and pseudospin degeneracy.
Introducing the characteristic length $\ell_{T}=\ell_0(k_BT\ell_0/\hbar v_g^{u})$,
this condition of smoothness can be rewritten as $\ell_{0} \ll \ell_{T}$,
where $v_{g}^{u}=\ell_{0}^{2}\, \hbar^{-1} dV(y_{r}^{u})/dy$ is the group velocity at the edge $y_{r}^{u}$.
Notice, for conditions of Fig. 1 it follows that $v_{g}^{u}/v_{F}=(\ell_{0}/\sqrt{2 \pi} \Delta y) \lll 1$,
due to $\ell_{0}/\Delta y \ll 1$. Point out, Eq. (\ref{eq18}) shows that at $y=y_{r}^{u}$ the Hall conductivity
changes its sign from the electron type of charge carriers to the hole one.

For $(\ell_T/\Delta y)^2\ll 1$,  Eq. (\ref{eq18}) can be rewritten as\cite{balev2011}
\begin{equation}
\sigma_{yx}^{(0)}(y)={2e^2\over h} \tanh\left(\frac{y_r^u-y}{2\ell_{T}}\right).
\label{eq21}
\end{equation}
 From Eq. (\ref{eq21})
it follows  \cite{bal2000,balev2011}
\begin{equation}
\frac{d\sigma_{yx}^{(0)}(y)}{dy}=
-{4e^2\over h} \left[\frac{1}{4\ell_T} cosh^{-2}\left(\frac{y-y_{r}^{u}}{2\ell_{T}}\right)\right].
\label{eq22}
\end{equation}
Further, for $L_{y}/2 \geq y \geq L_{y}/2-5\ell_{0}$ pertinent
numerical results \cite{brey,milt,gus,aba} for $\nu=2$ we approximate by the same analitical expression for
$\left[n(y)-p(y)\right]$ as in Ref. \onlinecite{balev2011}. That gives
\begin{equation}
\hspace*{-0.2cm}
\sigma_{yx}^{(0)}(y)=\frac{2e^{2}}{h} \int_{-\infty}^{\infty} \frac{dy_0}{\sqrt{\pi} \ell_{0}}
\,e^{-(y-y_0)^2/ \ell_0^2} \left[f_{0,y_0,-}-1\right] ,
\label{eq23}
\end{equation}
where it is used that $E_{0,y_0,-}$
is a sharply decreasing function at $y_{0} \approx y_{r}^{d}$ such that
the Fermi function in Eq. (\ref{eq23}) is very fastly growing at $y_{0} \approx y_{r}^{d}$ on a scale
$\ell_{d} \ll \ell_{0}$. Here appears a new characteristic scale $\ell_{d}=
k_{B}T \ell_{0}^{2}/\hbar |v_{g}^{d}|$ as
for a change of $y_{0}$ on $\ell_{d}$, at
$y_{0} \approx y_{r}^{d}$, the value of $E_{0,y_0,-}$ will change on $k_{B}T$;
point out that $|v_{g}^{d}| \ll v_{F}$ is implicit in Fig. 1.

From Eq. (\ref{eq23}) it follows \cite{balev2011} 
\begin{equation}
d\sigma_{yx}^{(0)}(y)/dy=(2e^2 /h\sqrt{\pi}\ell_0)\, e^{-(y-y_r^d)^2/\ell_0^2} ,
\label{eq24}
\end{equation}
by changing the derivatives over $y$ to those
over $y_{0}$ and integrating by parts.

In a similar manner, for  case (i) and  $y>0$, we obtain
from Eqs. (\ref{eq3})-(\ref{eq5}), (\ref{eq16}) that
$\sigma_{yx}^{(1)}({\bf r})=\sigma_{yx}^{(1)}(y)\cos \left(G x \right)$, where
\begin{eqnarray}
\nonumber
\sigma_{yx}^{(1)}(y)&=&-\frac{e^{2}}{h}  \frac{V_{s}}{k_{B} T}
\left[ \cosh^{-2}\left(\frac{y-y_{r}^{u}}{2\ell_{T}}\right) \right. \\
&&\left.+\frac{2 k_{B} T \ell_{0}}{\sqrt{\pi} \hbar |v_{g}^{d}|} e^{-\left(y-y_{r}^{d}\right)^{2}/\ell_{0}^{2}}   \right],
\label{eq25}
\end{eqnarray}
here it is taken into account that $v_{g}^{d}<0$.   Due to $k_{B}T \ell_{0}/\hbar |v_{g}^{d}| \ll 1$,
the amplitude of the
second term in the square brackets of Eq. (\ref{eq25}) is less than of the first one.
In addition, Eq. (\ref{eq25}) shows that a small parameter $V_{s}/k_{B}T \ll 1$ warrants
that $|\sigma_{yx}^{(1)}(x,y) \ll |\sigma_{yx}^{(0)}(y)|$; in particular, at $y \approx y_{r}^{u}$.

\section{Strong renormalization of the EMPs in graphene by a weak superlattice potential}

Now we will study effect of the superlattice potential, $V_{s}(x)$, on the fundamental EMPs for  case (i),
i.e., for conditions like in Fig. 1; dissipation is neglected. 
For $V_{s}(x) \equiv 0$, our fundamental EMPs
will coincide with ones obtained in Ref. \onlinecite{balev2011}. 
Similar with Ref. \onlinecite{balev2011}, we expect that the charge excitation due to EMPs at
the right part of channel will be strongly localized
at $y_{r}^{u}$ (  $\rho^{ru}(t,{\bf r})$) and $y_{r}^{d}$ ( $\rho^{rd}(t,{\bf r})$).
Then the components of the  current density ${\bf j}(\omega,{\bf r})$ in the
low-frequency limit $\omega\ll v_F/\ell_0$ are given, cf. with,\cite{balev2011} as
\begin{eqnarray}
\nonumber
\hspace*{-0.25cm}j_x(\omega,{\bf r})&=&-[\sigma_{yx}^{(0)}(y)+\sigma_{yx}^{(1)}({\bf r})]E_y(\omega,{\bf r}) \\*
&+&v_g^u \rho^{ru}(\omega,{\bf r})+v_g^d \rho^{rd}(\omega,{\bf r}) ,
\label{eq26}
\end{eqnarray}
\begin{equation}
j_y(\omega,{\bf r})=[\sigma_{yx}^{(0)}(y)+\sigma_{yx}^{(1)}({\bf r})]E_x(\omega,{\bf r}),
\label{eq27}
\end{equation}
where we have suppressed the factor $\exp(-i\omega t)$ common to all terms in Eqs. (\ref{eq26}) and (\ref{eq27}).
From  Eqs. (\ref{eq26})-(\ref{eq27}), Poisson's equation, the linearized continuity equation
and using that for EMPs ${\bf E}(\omega,{\bf r})=-{\bf \nabla} \varphi(\omega,{\bf r})$), we obtain
%
\begin{eqnarray}
\nonumber
\hspace*{-0.99cm}&&-i\omega(\rho^{ru}(\omega, {\bf r})+\rho^{rd}(\omega, {\bf r}))+
[v_g^u \partial_{x}\rho^{ru}(\omega, {\bf r}) \\*
\nonumber
&&+v_g^d \partial_{x} \rho^{rd}(\omega,{\bf r})]+
[\partial_{x} \sigma_{yx}^{(1)}({\bf r})] \partial_{y} \varphi(\omega,{\bf r}) \\*
&&-[\partial_{y} \sigma_{yx}^{(0)}(y)+\partial_{y} \sigma_{yx}^{(1)}({\bf r})]
\partial_{x} \varphi(\omega,{\bf r})=0 ,
\label{18d}
\end{eqnarray}
where $\partial_{x}=\partial/\partial x$. Point out, in Eq. (\ref{18d}) the coefficients are invariant
for translations along $x$ on any distance integral of the period $a_{0}$.
Then in Eq. (\ref{18d}) we assume that

\begin{eqnarray}
\nonumber
&&\rho^{ru}(\omega, {\bf r})=\sum_{\ell=-1}^{1}  \rho^{ru}_{\ell}(\omega,k_{x},y) e^{ik_{x}^{(\ell)} x} ,
\\*
\nonumber
&&\rho^{rd}(\omega, {\bf r})=\sum_{\ell=-1}^{1} \rho^{rd}_{\ell}(\omega,k_{x},y) e^{ik_{x}^{(\ell)} x} , \\*
&&\varphi(\omega,{\bf r})=\sum_{\ell=-1}^{1} \varphi_{\ell}(\omega,k_{x},y) e^{ik_{x}^{(\ell)} x}  ,
\label{19d}
\end{eqnarray}
where $\ell=-1,0, 1$, $k_{x}^{(\ell)}=k_{x}+2\pi \ell/a_{0}$,  and
$k_{x} \equiv k_{x}^{(0)}$.

In Eq. (\ref{19d}), for a metallic gate  at a distance $d$ from the 2DES 
(e.g., it can be a heavily doped Si separated from the graphene sheet by a SiO$_{2}$ layer of  thickness $d=300$ nm),
is given by
\begin{eqnarray}
\nonumber
&&\varphi_{\ell}(\omega,k_{x},y)=\frac{2}{\epsilon}
\int_{-\infty}^{\infty} dy^{\prime}
R_{g}(|y-y^{\prime}|,k_{x}^{(\ell)};d) \\*
&& \times \left[\rho^{ru}_{\ell}(\omega, k_x,y^{\prime})+\rho^{rd}_{\ell}(\omega, k_x,y^{\prime})\right] ,
\label{20d}
\end{eqnarray}
where $R_g(...)$ is given by
\begin{eqnarray}
\hspace*{-0.89cm} R_{g}(|y-y^{\prime}|,k_{x}^{(\ell)};d)&=&K_{0}(|k_{x}^{(\ell)}||y-y^{\prime }|) \nonumber \\
&-& K_{0}(|k_{x}^{(\ell)}|\sqrt{(y-y^{\prime })^{2}+4d^{2}}),
\label{21d}
\end{eqnarray}
where $K_0(x)$ is the modified Bessel function. Without of a metallic gate, $d \to \infty$, the
dielectric constant $\epsilon $ is spatially homogeneous if not stated otherwise. 

Multiplying Eq. (\ref{18d}) by $ i \exp(-ik_{x} x)$ and then integrating over $x$, $\int_{0}^{L_{x}} dx$, we
obtain
\begin{eqnarray}
\nonumber
\hspace*{-0.99cm}
&&(\omega-k_{x} v_g^u) \rho^{ru}_{0}(\omega,k_{x},y)+
(\omega-k_{x} v_g^d) \rho^{rd}_{0}(\omega,k_{x},y)  \\*
\nonumber
&&-\frac{\pi}{a_{0}} \sigma_{yx}^{(1)}(y)
\left[\partial_{y} \varphi_{-1}(\omega,k_{x},y)-\partial_{y} \varphi_{1}(\omega,k_{x},y) \right]\\*
\nonumber
&&+[\partial_{y} \sigma_{yx}^{(0)}(y)] k_{x} \varphi_{0}(\omega,k_{x},y)+\frac{1}{2}  [\partial_{y} \sigma_{yx}^{(1)}(y)]   \\*
&&\times \left[k_{x}^{(-1)}\varphi_{-1}(\omega,k_{x},y)+k_{x}^{(1)}\varphi_{1}(\omega,k_{x},y)  \right]=0.
\label{22d}
\end{eqnarray}
Further, multiplying Eq. (\ref{18d}) by $ i \exp(-ik_{x}^{(-1)} x)$ and then integrating
over $x$, $\int_{0}^{L_{x}} dx$, we obtain
\begin{eqnarray}
\nonumber
\hspace*{-0.99cm}
&&(\omega-k_{x}^{(-1)} v_g^u) \rho^{ru}_{-1}(\omega,k_{x},y)+
(\omega-k_{x}^{(-1)} v_g^d) \rho^{rd}_{-1}(\omega,k_{x},y)  \\*
\nonumber
&&+\frac{\pi}{a_{0}} \sigma_{yx}^{(1)}(y)\partial_{y} \varphi_{0}(\omega,k_{x},y)+
[\partial_{y} \sigma_{yx}^{(0)}(y)] k_{x}^{(-1)} \varphi_{-1}(\omega,k_{x},y)  \\*
&&+\frac{1}{2} k_{x} [\partial_{y} \sigma_{yx}^{(1)}(y)] \varphi_{0}(\omega,k_{x},y)=0.
\label{23d}
\end{eqnarray}
In addition, multiplying Eq. (\ref{18d}) by $ i \exp(-ik_{x}^{(1)} x)$ and then integrating
over $x$, $\int_{0}^{L_{x}} dx$, we obtain
\begin{eqnarray}
\nonumber
\hspace*{-0.99cm}
&&(\omega-k_{x}^{(1)} v_g^u) \rho^{ru}_{1}(\omega,k_{x},y)+
(\omega-k_{x}^{(1)} v_g^d) \rho^{rd}_{1}(\omega,k_{x},y)  \\*
\nonumber
&&-\frac{\pi}{a_{0}} \sigma_{yx}^{(1)}(y)\partial_{y} \varphi_{0}(\omega,k_{x},y)+
[\partial_{y} \sigma_{yx}^{(0)}(y)] k_{x}^{(1)} \varphi_{1}(\omega,k_{x},y)  \\*
&&+\frac{1}{2} k_{x} [\partial_{y} \sigma_{yx}^{(1)}(y)] \varphi_{0}(\omega,k_{x},y)=0.
\label{24d}
\end{eqnarray}
For $y_r^d - y_r^u\gg \ell_T$, from Eqs. (\ref{22d})-(\ref{24d})
it follows that $\rho^{ru}_{\ell}(\omega, k_x,y)$ and $\rho^{rd}_{\ell}(\omega, k_x,y)$
can be well approximated (cf. Ref. \onlinecite{balev2011}) as
\begin{eqnarray}
\nonumber
&&\rho^{ru}_{\ell}(\omega, k_x,y)=
\Big[4\ell_T\cosh^2({y-y_r^u\over 2\ell_T})\Big]^{-1}\,\rho^{ru}_{\ell}(\omega, k_x) , \\*
&&\rho^{rd}_{\ell}(\omega, k_x,y)= (1/\sqrt{\pi}\ell_0) e^{-(y-y_r^d)^2/\ell_0^2} \,\rho^{rd}_{\ell}(\omega, k_x).
\label{25d}
\end{eqnarray}
In addition, we can neglect by overlap between $\rho^{ru}(\omega, k_x,y)$ and $\rho^{rd}(\omega, k_x,y)$
in Eqs. (\ref{22d})-(\ref{24d}). Then, by  integration of Eqs. (\ref{22d})-(\ref{24d}) over $y$
within separate regions around $y_{r}^{u}$ and $y_{r}^{d}$, we obtain by straightforward calculations
six coupled linear homogeneous equations for six unknown functions:
$\rho^{ru}_{\ell}(\omega, k_x)$ and
$\rho^{rd}_{\ell}(\omega, k_x)$, where $\ell=-1, 0, 1$. They read, with  $y_{r}^{du}\equiv y_{r}^{d}-y_{r}^{u}>0$,
\begin{eqnarray}
\nonumber
\hspace*{-0.99cm}
&&[\omega-\omega_{+,0}^{(i)}(k_{x};d)] \rho^{ru}_{0}(\omega,k_{x})-
2c_{h}k_{x}R_{g}(y_{r}^{du},k_{x};d) \\*
\nonumber
&&\times \rho^{rd}_{0}(\omega,k_{x}) +c_{h} k_{x} \left(\frac{V_{s}}{k_{B}T} \ell_{T}\right)
\left[R_{g}^{\prime}(y_{r}^{du},k_{x}^{(-1)};d) \right. \\*
&&\times  \left.\rho^{rd}_{-1}(\omega,k_{x}) +R_{g}^{\prime}(y_{r}^{du},k_{x}^{(1)};d) \rho^{rd}_{1}(\omega,k_{x})
\right]=0,
\label{26d}
\end{eqnarray}
\begin{eqnarray}
\nonumber
\hspace*{-0.99cm}
&&[\omega-\omega_{-,0}^{(i)}(k_{x};d)] \rho^{rd}_{0}(\omega,k_{x})+
c_{h}k_{x}R_{g}(y_{r}^{du},k_{x};d)  \\*
\nonumber
&&\times \rho^{ru}_{0}(\omega,k_{x})-c_{h} k_{x} \left(\frac{V_{s}}{k_{B}T} \ell_{T} \frac{v_{g}^{u}}{2|v_{g}^{d}|}\right)
\left[R_{g}^{\prime}(y_{r}^{du},k_{x}^{(-1)};d)\right. \\*
&&\left. \times  \rho^{ru}_{-1}(\omega,k_{x}) +R_{g}^{\prime}(y_{r}^{du},k_{x}^{(1)};d) \rho^{ru}_{1}(\omega,k_{x})
\right]=0,
\label{27d}
\end{eqnarray}
\begin{eqnarray}
\nonumber
\hspace*{-0.99cm}
&&[\omega-\omega_{+,0}^{(i)}(k_{x}^{(-1)};d)] \rho^{ru}_{-1}(\omega,k_{x})-
2c_{h}k_{x}^{(-1)} \\*
\nonumber
&&\times R_{g}(y_{r}^{du},k_{x}^{(-1)};d) \rho^{rd}_{-1}(\omega,k_{x}) +c_{h} k_{x}^{(-1)} \left(\frac{V_{s}}{k_{B}T} \ell_{T}\right)
 \\*
&&\times R_{g}^{\prime}(y_{r}^{du},k_{x};d) \rho^{rd}_{0}(\omega,k_{x})=0,
\label{28d}
\end{eqnarray}
\begin{eqnarray}
\nonumber
\hspace*{-0.99cm}
&&[\omega-\omega_{-,0}^{(i)}(k_{x}^{(-1)};d)] \rho^{rd}_{-1}(\omega,k_{x})+
c_{h}k_{x}^{(-1)} \\*
\nonumber
&&\times R_{g}(y_{r}^{du},k_{x}^{(-1)};d) \rho^{ru}_{-1}(\omega,k_{x}) -c_{h} k_{x}^{(-1)}  \left(\frac{V_{s}}{k_{B}T} \right.
 \\*
&&\left. \times \ell_{T} \frac{v_{g}^{u}}{2|v_{g}^{d}|}\right)
 R_{g}^{\prime}(y_{r}^{du},k_{x};d)  \rho^{ru}_{0}(\omega,k_{x})=0,
\label{29d}
\end{eqnarray}
\begin{eqnarray}
\nonumber
\hspace*{-0.99cm}
&&[\omega-\omega_{+,0}^{(i)}(k_{x}^{(1)};d)] \rho^{ru}_{1}(\omega,k_{x})-
2c_{h}k_{x}^{(1)} \\*
\nonumber
&&\times R_{g}(y_{r}^{du},k_{x}^{(1)};d) \rho^{rd}_{1}(\omega,k_{x}) +c_{h} k_{x}^{(1)} \left(\frac{V_{s}}{k_{B}T} \ell_{T}\right) \\*
&&\times R_{g}^{\prime}(y_{r}^{du},k_{x};d) \rho^{rd}_{0}(\omega,k_{x})=0,
\label{30d}
\end{eqnarray}
\begin{figure}[ht]
\vspace*{-0.5cm}
\includegraphics [height=11cm, width=8cm]{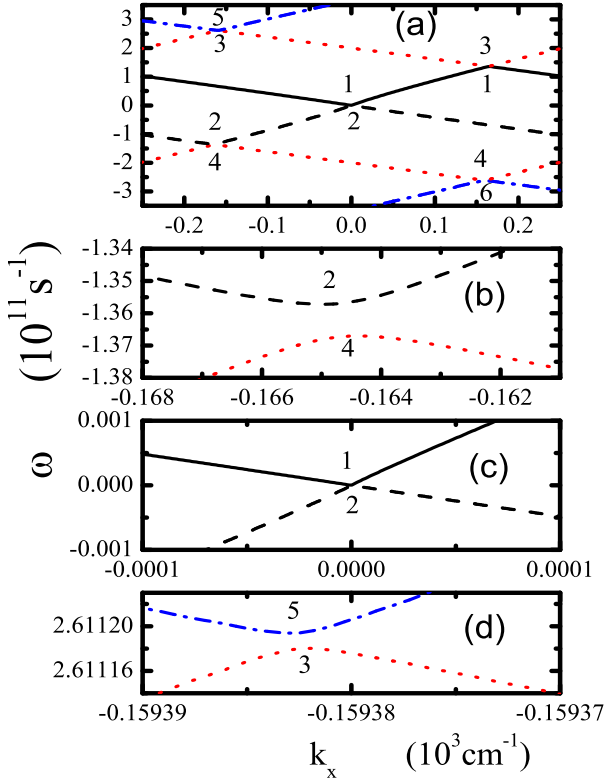}
\vspace*{-0.7cm}
\caption{(Color online) The dispersion relations $\omega(k_x, d\to \infty)$, without the gate, of six EMPs
calculated from Eqs. (\ref{26d})-(\ref{31d})
for $2\pi/a_{0}=500$\,cm$^{-1}$, $V_{s}/k_{B}T=0.3$,
$y_{r}^{du}=10 \ell_{0}$, $\Delta y=10 \ell_{0}$, $B=9$T, $T=77$K, $\ell_{T}/\ell_{0}=2$,  $v_{g}^{u}=4 \times 10^{6}$ cm/s,
$v_{g}^{d}=-3 \times 10^{7}$ cm/s, $\epsilon=2$.
Panel (a) presents the dispersion relations within the first Brillouin zone,
$\pi/a_{0}> k_{x} \geq -\pi/a_{0}$, by the curves 1 (solid), 2 (dashed), 3 and 4 (dotted),
5 and 6 (dash-dotted).
Panel (b) presents a zoom of the anticrossing for the branches 2 and 4,
at $k_{x} \approx -165$\,cm$^{-1}$ and $\omega \approx -1.36 \times 10^{11}$\,s$^{-1}$,
with the gap $\approx 0.98 \times 10^{9}$\,s$^{-1}$; here the EMPs 2 and 4 have a  zero value of group velocity for pertinent $k_{x}$.
Panel (c) presents a zoom of the branches 1 and 2
at $k_{x} \approx 0$ and $\omega \approx 0$; here a finite gap is absent.
Panel (d) presents a zoom of the anticrossing for the branches 3 and 5.
}
\end{figure}

\begin{figure}[ht]
\vspace*{-0.5cm}
\includegraphics [height=11cm, width=8cm]{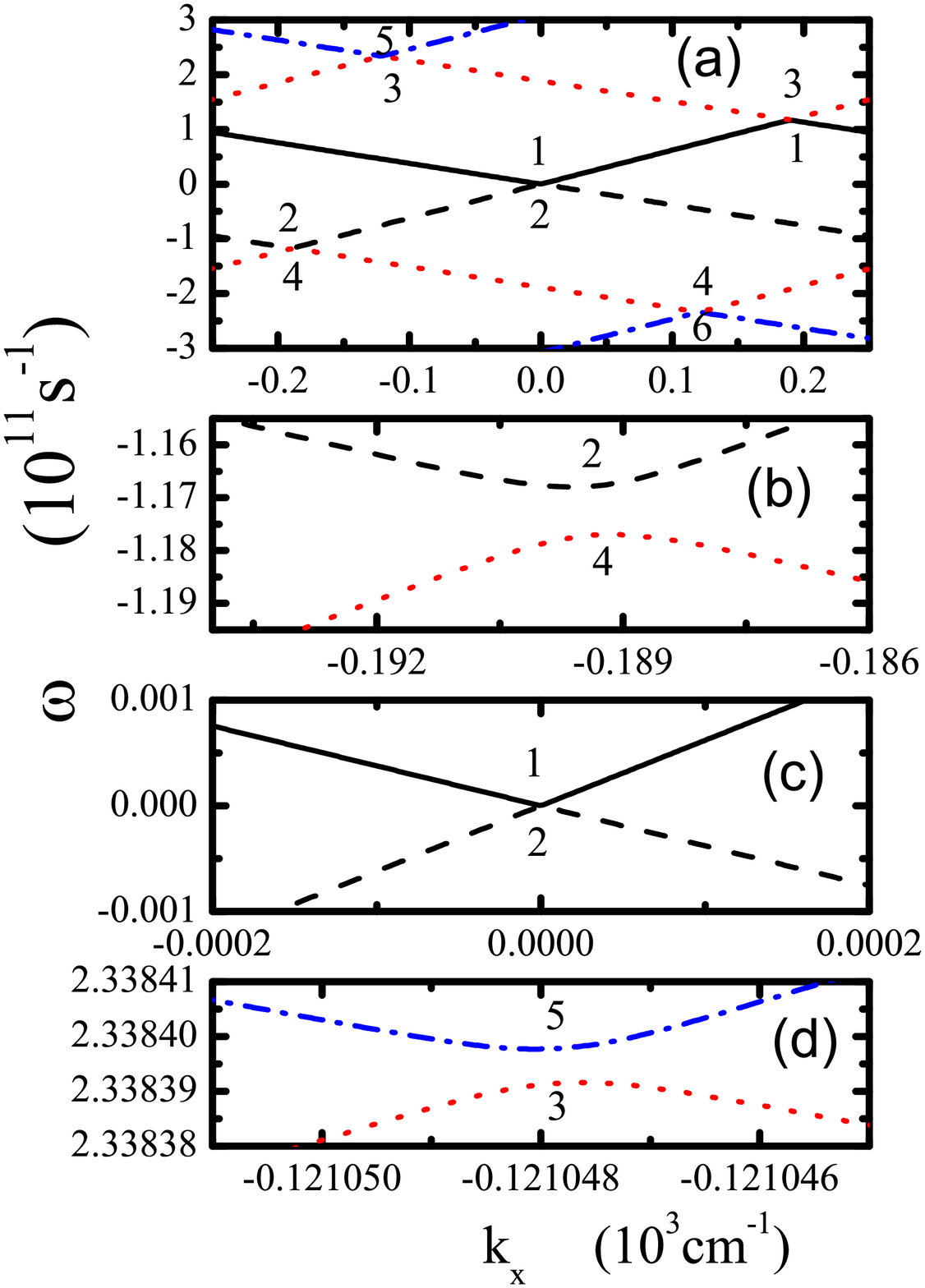}
\vspace*{-0.7cm}
\caption{(Color online) The dispersion relations $\omega(k_x, d=3000$nm) of six EMPs
calculated from Eqs. (\ref{26d})-(\ref{31d})
for $2\pi/a_{0}=500$\,cm$^{-1}$ and other parameters of Fig. 2 except $d=3000$nm.
Panel (a) presents the dispersion relations within the first Brillouin zone by the curves 1 (solid), 2 (dashed), 3 and 4 (dotted),
5 and 6 (dash-dotted).
Panel (b) presents a zoom of the anticrossing for the branches 2 and 4,
at $k_{x} \approx -189$\,cm$^{-1}$ and $\omega \approx -1.17 \times 10^{11}$\,s$^{-1}$,
with the gap $\approx 0.90 \times 10^{9}$\,s$^{-1}$.
Panel (c) presents a zoom of the branches 1 and 2
at $k_{x} \approx 0$ and $\omega \approx 0$.
Panel (d) presents a zoom of the anticrossing for the branches 3 and 5.
}
\end{figure}

\begin{figure}[ht]
\vspace*{-0.5cm}
\includegraphics [height=11cm, width=8cm]{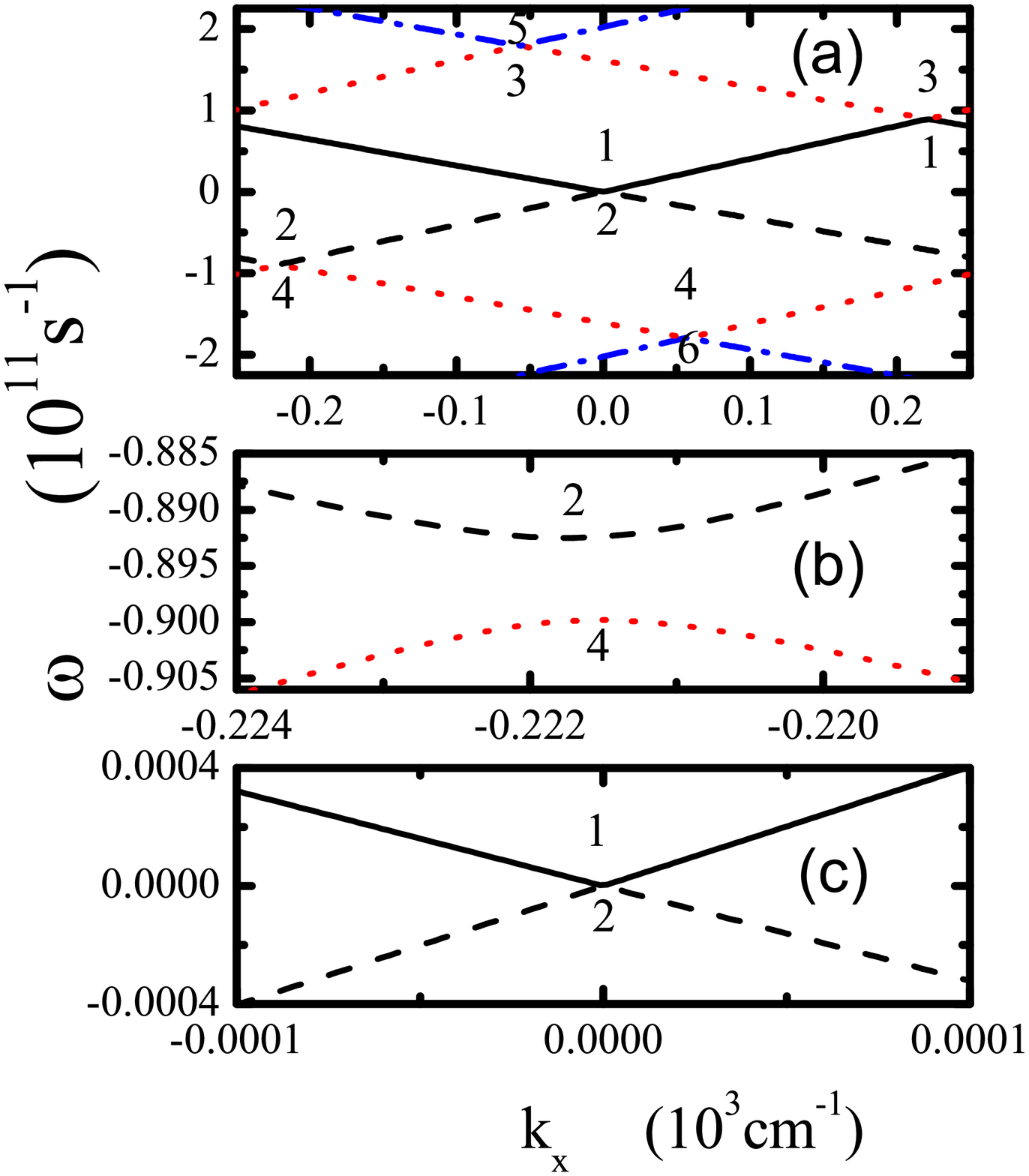}
\vspace*{-2.5cm}
\caption{(Color online) The dispersion relations $\omega(k_x, d=300$nm) of six EMPs
calculated from Eqs. (\ref{26d})-(\ref{31d})
for $2\pi/a_{0}=500$\,cm$^{-1}$ and other parameters of Fig. 2 except $d=300$nm.
Panel (a) presents the dispersion relations within the first Brillouin zone by the curves 1 (solid), 2 (dashed), 3 and 4 (dotted),
5 and 6 (dash-dotted).
Panel (b) presents a zoom of the anticrossing for the branches 2 and 4,
at $k_{x} \approx -222$\,cm$^{-1}$ and $\omega \approx -0.90 \times 10^{11}$\,s$^{-1}$,
with the gap $\approx 0.73 \times 10^{9}$\,s$^{-1}$.
Panel (c) presents a zoom of the branches 1 and 2
at $k_{x} \approx 0$ and $\omega \approx 0$.}
\end{figure}

\begin{eqnarray}
\nonumber
\hspace*{-0.99cm}
&&[\omega-\omega_{-,0}^{(i)}(k_{x}^{(1)};d)] \rho^{rd}_{1}(\omega,k_{x})+
c_{h}k_{x}^{(1)} \\*
\nonumber
&&\times R_{g}(y_{r}^{du},k_{x}^{(1)};d) \rho^{ru}_{1}(\omega,k_{x}) -c_{h} k_{x}^{(1)}
\left(\frac{V_{s}}{k_{B}T} \right.  \\*
&&\times \left. \ell_{T} \frac{v_{g}^{u}}{2|v_{g}^{d}|}\right) R_{g}^{\prime}(y_{r}^{du},k_{x};d) \rho^{ru}_{0}(\omega,k_{x})=0,
\label{31d}
\end{eqnarray}
\begin{figure}[ht]
\vspace*{-0.5cm}
\includegraphics [height=11cm, width=8cm]{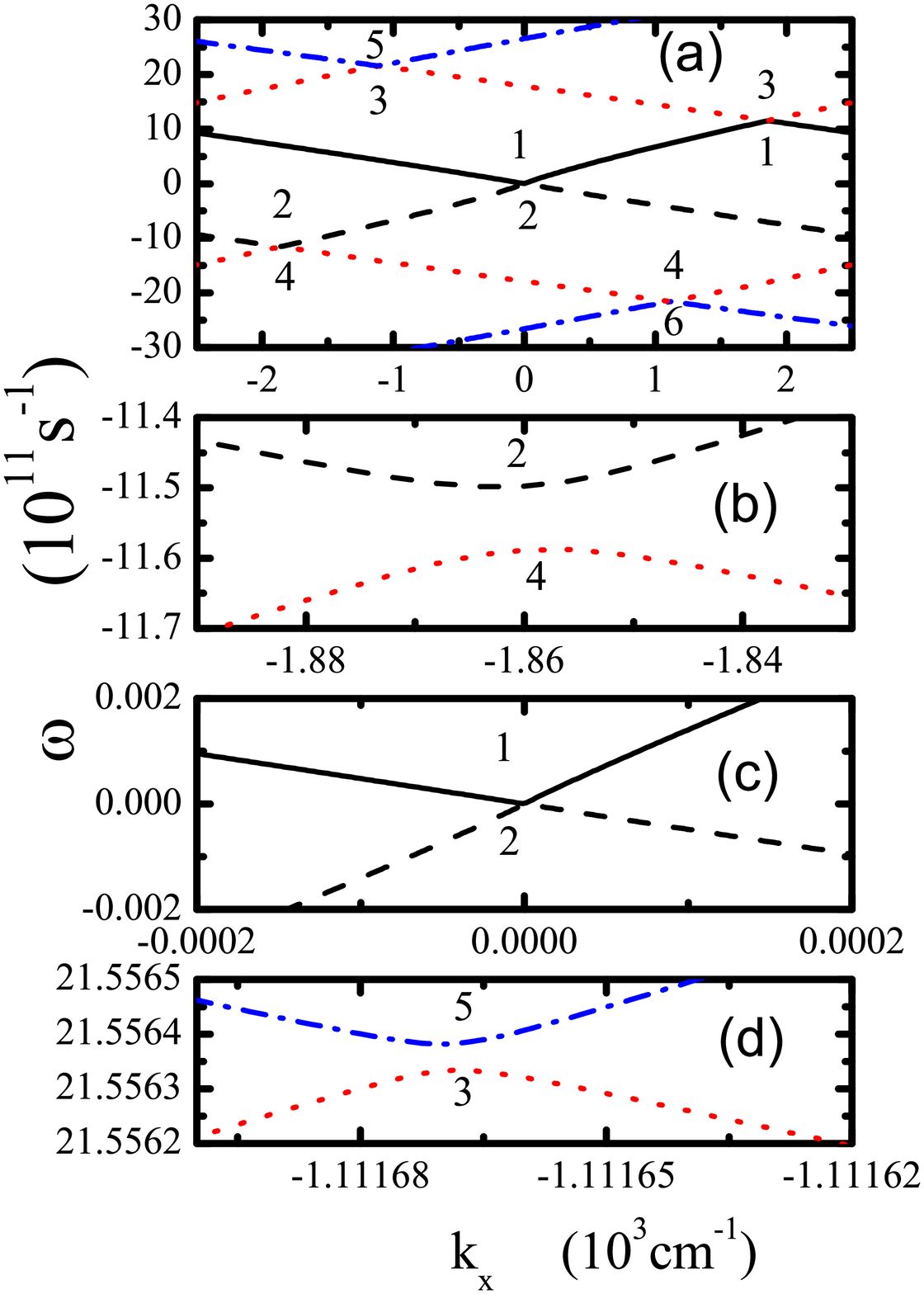}
\vspace*{-0.7cm}
\caption{(Color online) The dispersion relations $\omega(k_x, d\to \infty)$, without the gate, of six EMPs
calculated from Eqs. (\ref{26d})-(\ref{31d})
for $2\pi/a_{0}=5000$\,cm$^{-1}$, the rest of parameters are the same as in Fig. 2.
Panel (a) presents the dispersion relations within the first Brillouin zone,
$\pi/a_{0}> k_{x} \geq -\pi/a_{0}$, by the curves 1 (solid), 2 (dashed), 3 and 4 (dotted),
5 and 6 (dash-dotted).
Panel (b) presents a zoom of the anticrossing for the branches 2 and 4,
at $k_{x} \approx -1860$\,cm$^{-1}$ and $\omega
\approx -1.15 \times 10^{12}$\,s$^{-1}$, 
with the gap $\approx 9.0 \times 10^{9}$\,s$^{-1}$.
Panel (c) presents a zoom of the branches 1 and 2
at $k_{x} \approx 0$ and $\omega \approx 0$; here a finite gap is absent.
Panel (d) presents a zoom of the anticrossing for the branches 3 and 5.
}
\end{figure}

\begin{figure}[ht]
\vspace*{-0.5cm}
\includegraphics [height=11cm, width=8cm]{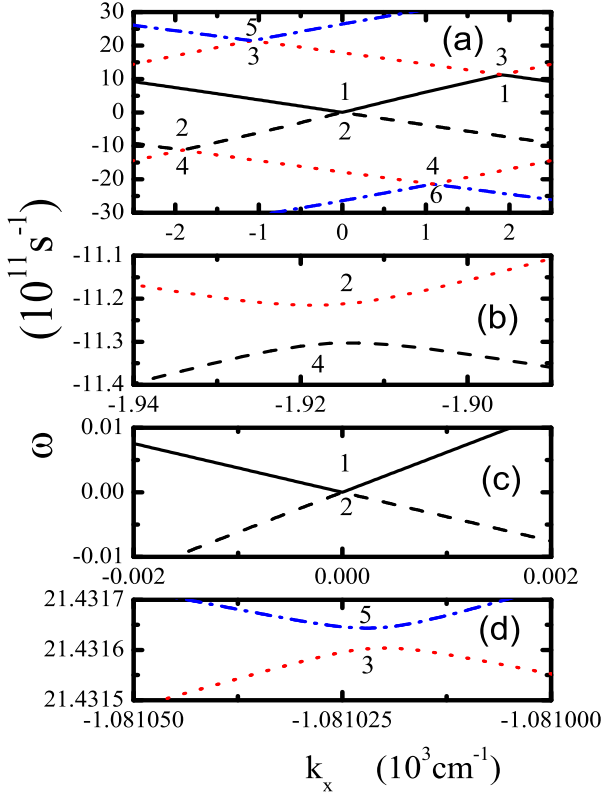}
\vspace*{-0.7cm}
\caption{(Color online) The dispersion relations $\omega(k_x, d=3000$nm) of six EMPs
calculated from Eqs. (\ref{26d})-(\ref{31d})
for the gate at $d=3000$nm; as $2\pi/a_{0}=5000$\,cm$^{-1}$ so the other
parameters are the same as in Fig. 5.
Panel (a) presents the dispersion relations within the first Brillouin zone.
Panel (b) presents a zoom of the anticrossing for the branches 2 and 4,
at $k_{x} \approx -1920$\,cm$^{-1}$ and $\omega \approx -1.13
\times 10^{12}$\,s$^{-1}$, with the gap
with the gap $\approx 8.8 \times 10^{9}$\,s$^{-1}$.
Panel (c) presents a zoom of the branches 1 and 2
at $k_{x} \approx 0$ and $\omega \approx 0$; here a finite gap is absent.
Panel (d) presents a zoom of the anticrossing for the branches 3 and 5.}
\end{figure}

\begin{figure}[ht]
\vspace*{-0.5cm}
\includegraphics [height=11cm, width=8cm]{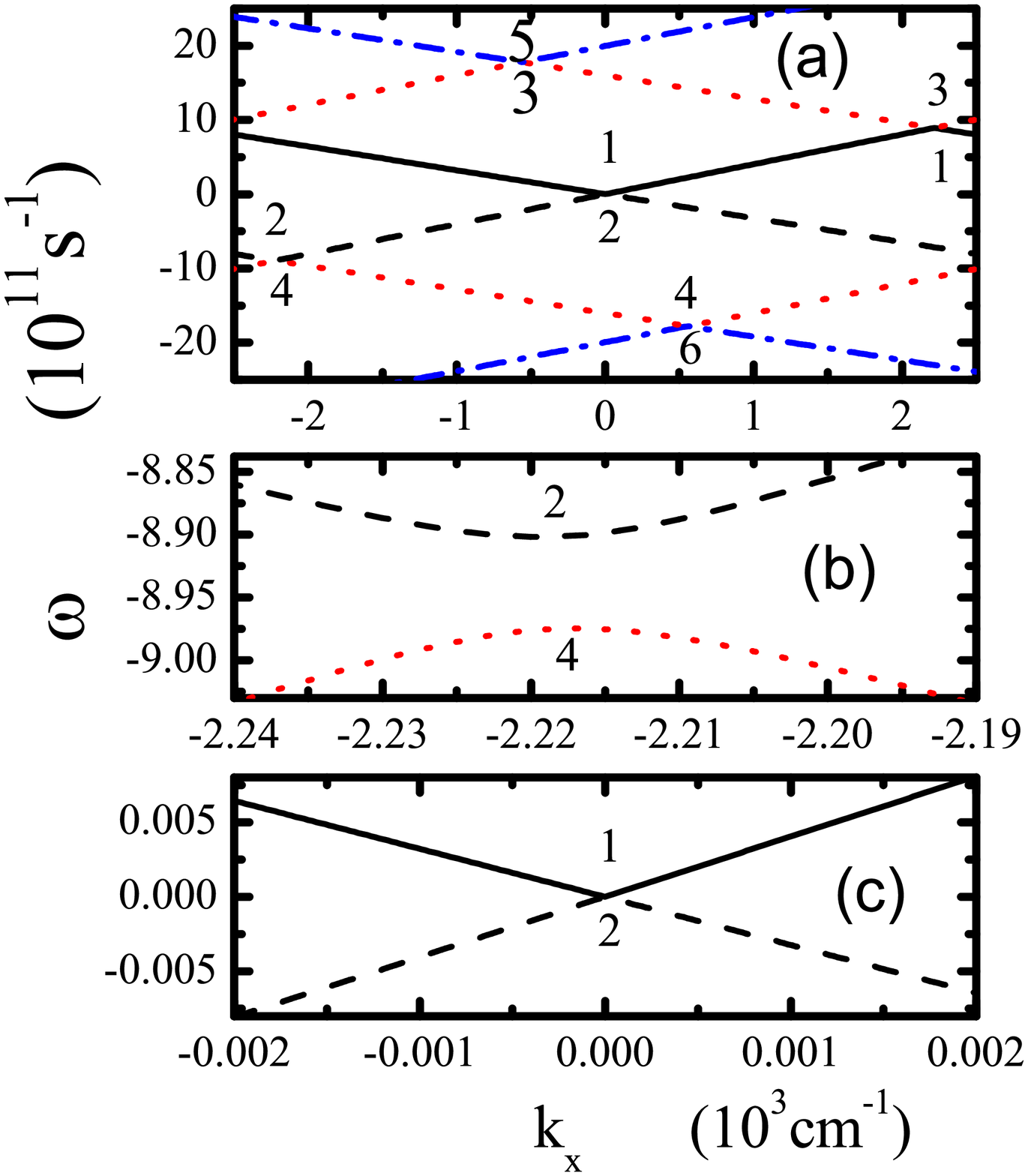}
\vspace*{-2.5cm}
\caption{(Color online) The dispersion relations $\omega(k_x, d=300$nm) of six EMPs
calculated from Eqs. (\ref{26d})-(\ref{31d})
for $2\pi/a_{0}=5000$\,cm$^{-1}$ and other parameters of Fig. 5 except $d=300$nm.
Panel (a) presents the dispersion relations within the first Brillouin zone.
Panel (b) presents a zoom of the anticrossing for the branches 2 and 4,
at $k_{x} \approx -2220$\,cm$^{-1}$ and $\omega \approx -8.95 \times 10^{11}$\,s$^{-1}$, 
with the gap $\approx 7.3 \times 10^{9}$\,s$^{-1}$.
Panel (c) presents a zoom of the branches 1 and 2
at $k_{x} \approx 0$ and $\omega \approx 0$; here a finite gap is absent.}
\end{figure}
\begin{figure}[ht]
\vspace*{-0.5cm}
\includegraphics [height=11cm, width=8cm]{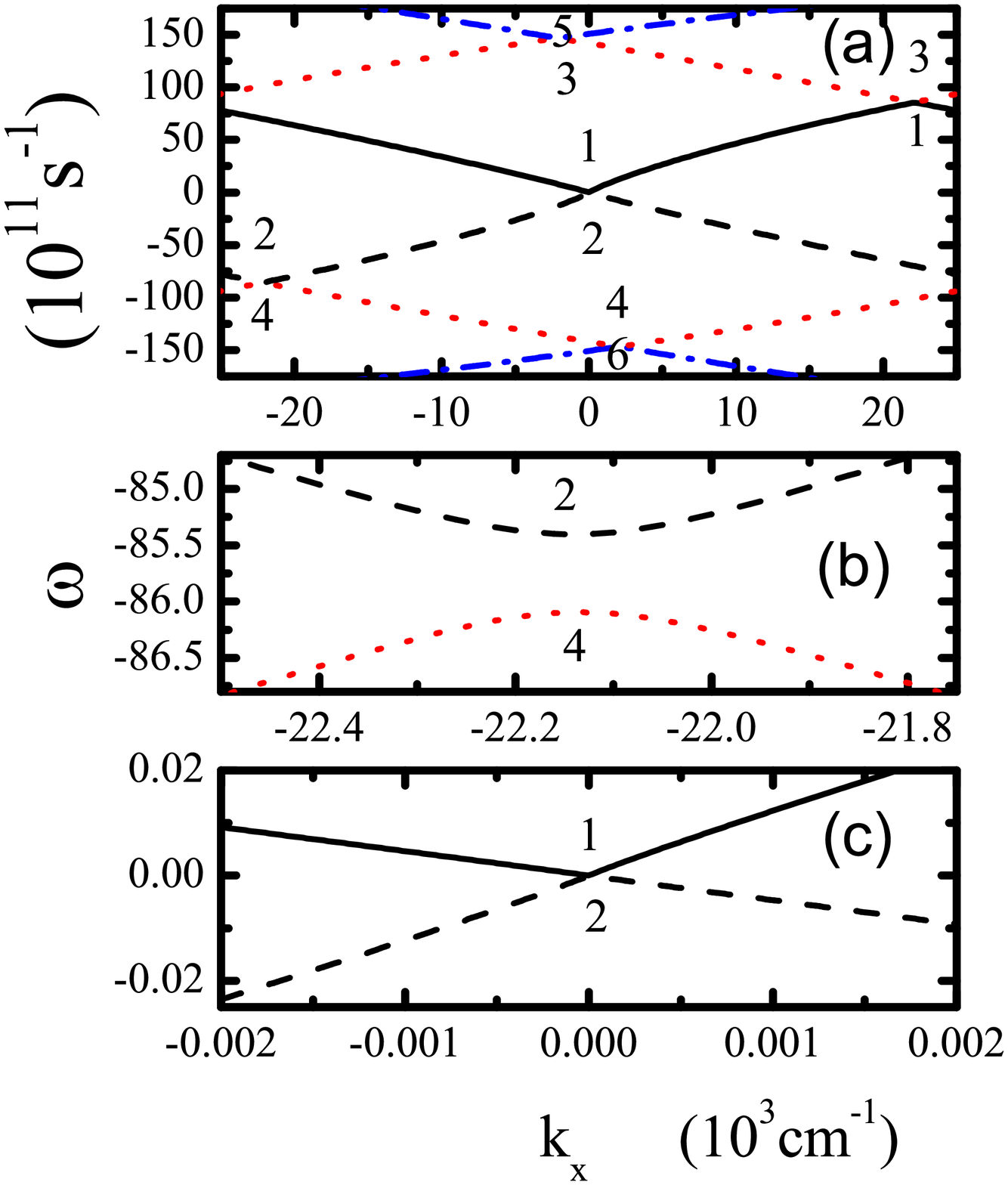}
\vspace*{-2.5cm}
\caption{(Color online) The dispersion relations $\omega(k_x, d\to \infty)$, without the gate, of six EMPs
calculated from Eqs. (\ref{26d})-(\ref{31d})
for $2\pi/a_{0}=50000$\,cm$^{-1}$, the rest of parameters are the same as in Fig. 2.
Panel (a) presents the dispersion relations within the first Brillouin zone,
$\pi/a_{0}> k_{x} \geq -\pi/a_{0}$, by the curves 1 (solid), 2 (dashed), 3 and 4 (dotted),
5 and 6 (dash-dotted).
Panel (b) presents a zoom of the anticrossing for the branches 2 and 4, at
$k_{x} \approx -22150$\,cm$^{-1}$ and $\omega \approx -8.6 \times 10^{12}$\,s$^{-1}$,
with the gap $\approx 6.9 \times 10^{10}$\,s$^{-1}$.
Panel (c) presents a zoom of the branches 1 and 2
at $k_{x} \approx 0$ and $\omega \approx 0$; a finite gap is absent.}
\end{figure}
where, by using the same notations as in Ref. \onlinecite{balev2011}, we have $c_{h}=4e^{2}/h \epsilon$,
\begin{equation}
\omega_{+,0}^{(i)}(k_x,d)=k_x v_g^u+ 2c_{h}k_x  a_{p}(k_{x};d) ,
\label{32d}
\end{equation}
\begin{figure}[ht]
\vspace*{-0.5cm}
\includegraphics [height=11cm, width=8cm]{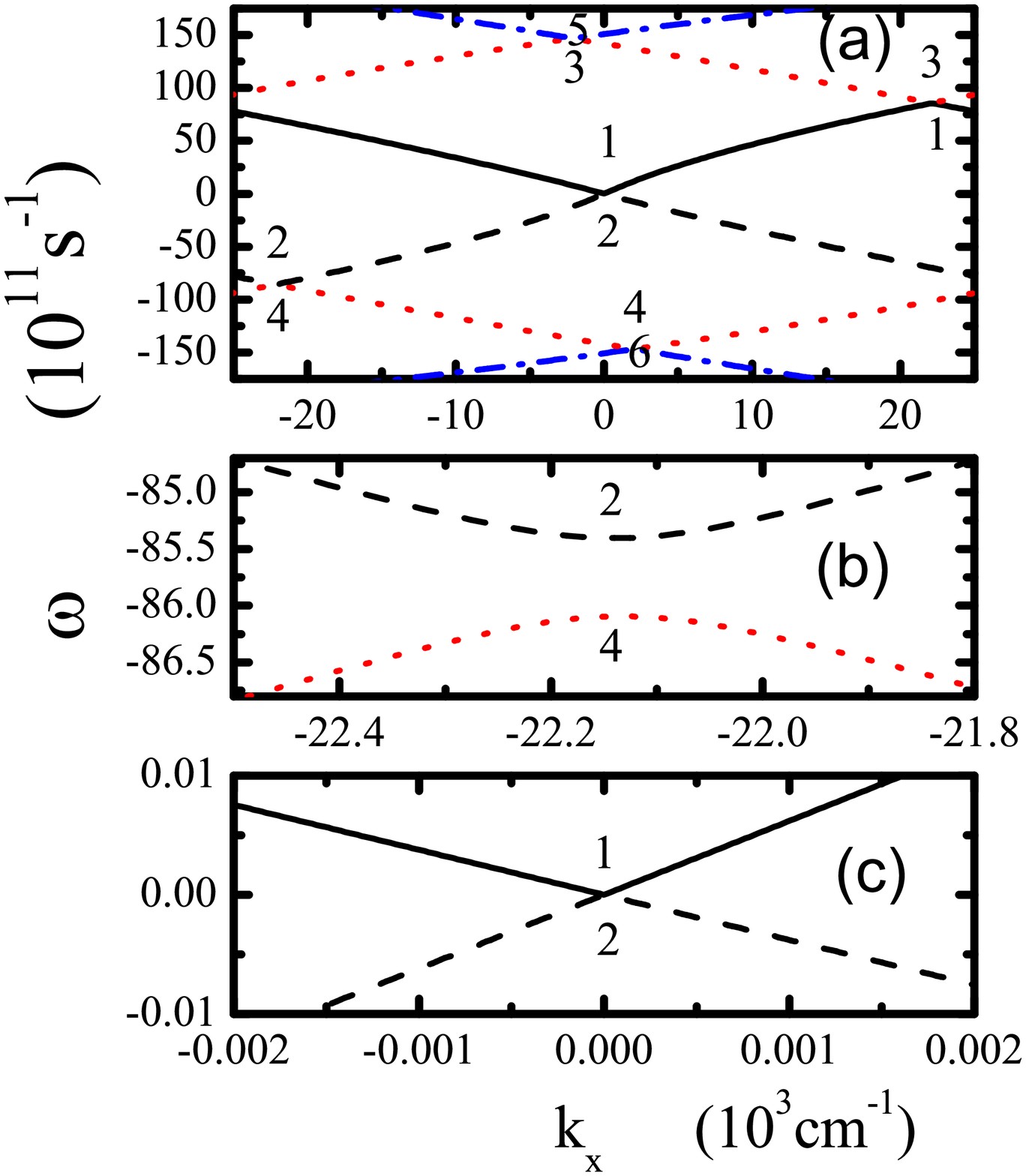}
\vspace*{-2.5cm}
\caption{(Color online) The dispersion relations $\omega(k_x, d=3000$nm)
of the EMPs calculated from Eqs. (\ref{26d})-(\ref{31d})
for 
parameters of Fig. 8, except $d=3000$nm.
Panel (a) presents the dispersion relations within the first Brillouin zone.
Panel (b) presents a zoom of the anticrossing for the branches 2 and 4,
with the gap $\approx 6.9 \times 10^{10}$\,s$^{-1}$.
Panel (c) presents a zoom of the branches 1 and 2
at $k_{x} \approx 0$ and $\omega \approx 0$; a finite gap is absent.
}
\end{figure}
\begin{equation}
\omega_{-,0}^{(i)}(k_x,d)=-k_x|v_g^d|-c_{h}k_x  \, a_{m}(k_{x};d) ,
\label{33d}
\end{equation}
with the matrix elements 
%
\begin{equation}
\hspace*{-0.3cm}a_{p}(k_{x};d)=\frac{1}{16} \int_{-\infty}^{\infty} \int_{-\infty}^{\infty}
\frac{dx dt \; R_{g}(\ell_{T}|x-t|,k_{x};d)}{\cosh^{2}(x/2)\cosh^{2}(t/2)}  ,
\label{34d}
\end{equation}
\begin{equation}
a_{m}(k_{x};d)=\frac{1}{\pi} \int_{-\infty}^{\infty} \int_{-\infty}^{\infty} \frac{dx dt}{e^{x^2+t^2}}
R_{g}(\ell_{0}|x-t|,k_{x};d) .
\label{35d}
\end{equation}
Point out, if the graphene is located between two dielectric media with dielectric constants: $\epsilon_{1}$ for the halfspace
below of the graphene, and $\epsilon_{2}$ for the dielectric layer between the graphene and a metallic gate, - then
in all above expressions $\epsilon=(\epsilon_{1}+\epsilon_{2})/2$. 

In addition, in Eqs. (\ref{26d})-(\ref{31d})
\begin{eqnarray}
\hspace*{-0.89cm} && R_{g}^{\prime}(y_{r}^{d}-y_{r}^{u},k_{x};d)=|k_{x}| \left[K_{1}(|k_{x}|y_{r}^{du}) \right.
\nonumber \\*
&&\left.-\frac{y_{r}^{du}}{\sqrt{(y_{r}^{du})^{2}+4d^2}}
 K_{1}(|k_{x}|\sqrt{(y_{r}^{du})^{2}+4d^{2}})\right] .
\label{36d}
\end{eqnarray}
Notice that  in the long-wavelength limit, $k_{x} \ell_{T} \ll 1$, and for large $d$, such that the effect of
gate, $\propto \exp(-2|k_{x}| d) \ll 1$,  can be neglected, from Eqs. (\ref{21d}),
(\ref{32d})-(\ref{36d}) we obtain that:  $R_{g}(|y-y^{\prime}|,k_{x};d) \approx  \ln(2/|k_x(y-y')|)-\gamma$,
where $\gamma$ is the Euler constant,
$a_{p}(k_{x};d) \approx \left[\ln(1/|k_x|\ell_T)-0.145\right]$,
$a_{m}(k_{x};d) \approx \left[\ln(1/|k_x|\ell_0)+3/4\right]$,
and $R_{g}^{\prime}(y_{r}^{du},k_{x};d) \approx 1/y_{r}^{du}$.

For $\nu=2$ and  case (i), in Fig. 2 we plot the dispersion relations $\omega(k_x,d\to \infty)$ of the EMPs
calculated from Eqs. (\ref{26d})-(\ref{31d}) for $2\pi/a_{0}=500$\,cm$^{-1}$, $V_{s}/k_{B}T=0.3$,
$y_{r}^{du}=\Delta y=10 \ell_{0}$, $B=9$T, $T=77$K, $\ell_{T}/\ell_{0}=2$,  $v_{g}^{u}=4 \times 10^{6}$ cm/s,
$v_{g}^{d}=-3 \times 10^{7}$ cm/s, $\epsilon=2$, and $\ell_{0}\approx 8.5$ nm.
Fig. 2(a) presents the dispersion relations of these EMPs within the first Brillouin zone
by the curves 1-6. Notice,  Fig. 2(a) shows that the branches 3 and 4 have
$\omega \approx \pm 2.0 \times 10^{11}$\,s$^{-1}$ at $k_{x} \approx 0$.
Fig. 2(b) presents a zoom of the anticrossing for the branches 2 and 4,
at $k_{x} \approx -165$\,cm$^{-1}$ and $\omega \approx -1.36 \times 10^{11}$\,s$^{-1}$, with the gap
$\approx 0.98 \times 10^{9}$\,s$^{-1}$: here the EMPs 2 and 4 have a  zero value of group velocity for pertinent $k_{x}$.
A panel for the anticrossing of the branches 1 and 3 it follows from the Fig. 2(b) by changing $k_{x}$ on $-k_{x}$
and $\omega$ on $-\omega$.
Fig. 2(c) presents a zoom of the branches 1 and 2
at $k_{x} \approx 0$ and $\omega \approx 0$; here a finite gap is absent.
Panel (d) presents a zoom of the anticrossing for the branches 3 and 5
at $k_{x} \approx -159.4$\,cm$^{-1}$ and $\omega \approx 2.611 \times 10^{11}$\,s$^{-1}$, with the gap
$\approx 1.37 \times 10^{6}$\,s$^{-1}$.

In Fig. 2 and below, it is used that the exact dispersion relation $\omega(k_x,d)$
of any EMP mode can be presented in the form periodic in the reciprocal space, i.e, $\omega(k_x,d)=\omega(k_x \pm 2\pi/a_{0},d)$,
and continuous across the borders of the Brillouin zone, $\omega(\pi/a_{0}-0,d)=\omega(\pi/a_{0}+0,d)$. The latter, in particular, does not allow
an infinite group velocity for the EMP. Point out that the dispersion curves  1 (solid), 2 (dashed), 3 and 4 (dotted) have correct periodic and continuous form
in the reciprocal space, $k_{x}$, and both qualitatively and quantitavely well describe dispersion of pertinent EMP modes in graphene with the superlattice:
1 and 2 are the main fundamental EMPs, 3 and 4 are the first excited fundamental EMPs. 
However, an approximate dispersion curves 5 and 6 (dash-dotted) qualitatively
correctly represent pertinent exact dependencies only nearby the anticrossings of 5 with 3 and of 6 with 4.
So the curves 5 and 6 are shown only within a small part of the first Brillouin zone in Fig. 2(a). In addition, as the second 
order contributions over the periodic
potential are neglected (as well as an additional contributions in Eqs. (19) with the $\ell=\pm 2$) the Fig. 2(d) gives
 only rough approximation for this anticrossing and, in particular, for its gap.

In Fig. 3 we plot the dispersion relations $\omega(k_x,d=3000$nm) of the EMPs
calculated from Eqs. (\ref{26d})-(\ref{31d}) for $d=3000nm$, the rest of parameters are the same as in Fig. 2.
Fig. 3(a) shows that the branches 3 and 4 have
$\omega \approx \pm 1.9 \times 10^{11}$\,s$^{-1}$ at $k_{x} \approx 0$.
Fig. 3(b) presents a zoom of the anticrossing for the branches 2 and 4,
at $k_{x} \approx -189$\,cm$^{-1}$ and $\omega \approx -1.17 \times 10^{11}$\,s$^{-1}$, with the gap
$\approx 0.90 \times 10^{9}$\,s$^{-1}$: here the EMPs 2 and 4 have a  zero value of group velocity for pertinent $k_{x}$.
A panel for the anticrossing of the branches 1 and 3 it follows from the Fig. 3(b) by changing $k_{x}$ on $-k_{x}$
and $\omega$ on $-\omega$.
Fig. 3(c) presents a zoom of the branches 1 and 2
at $k_{x} \approx 0$ and $\omega \approx 0$; here a finite gap is absent.
Fig. 3(d) presents a zoom of the anticrossing for the branches 3 and 5
at $k_{x} \approx -121$\,cm$^{-1}$ and $\omega \approx 2.34 \times 10^{11}$\,s$^{-1}$, with the gap
$\sim 5 \times 10^{5}$\,s$^{-1}$.

\begin{figure}[ht]
\vspace*{-0.5cm}
\includegraphics [height=11cm, width=8cm]{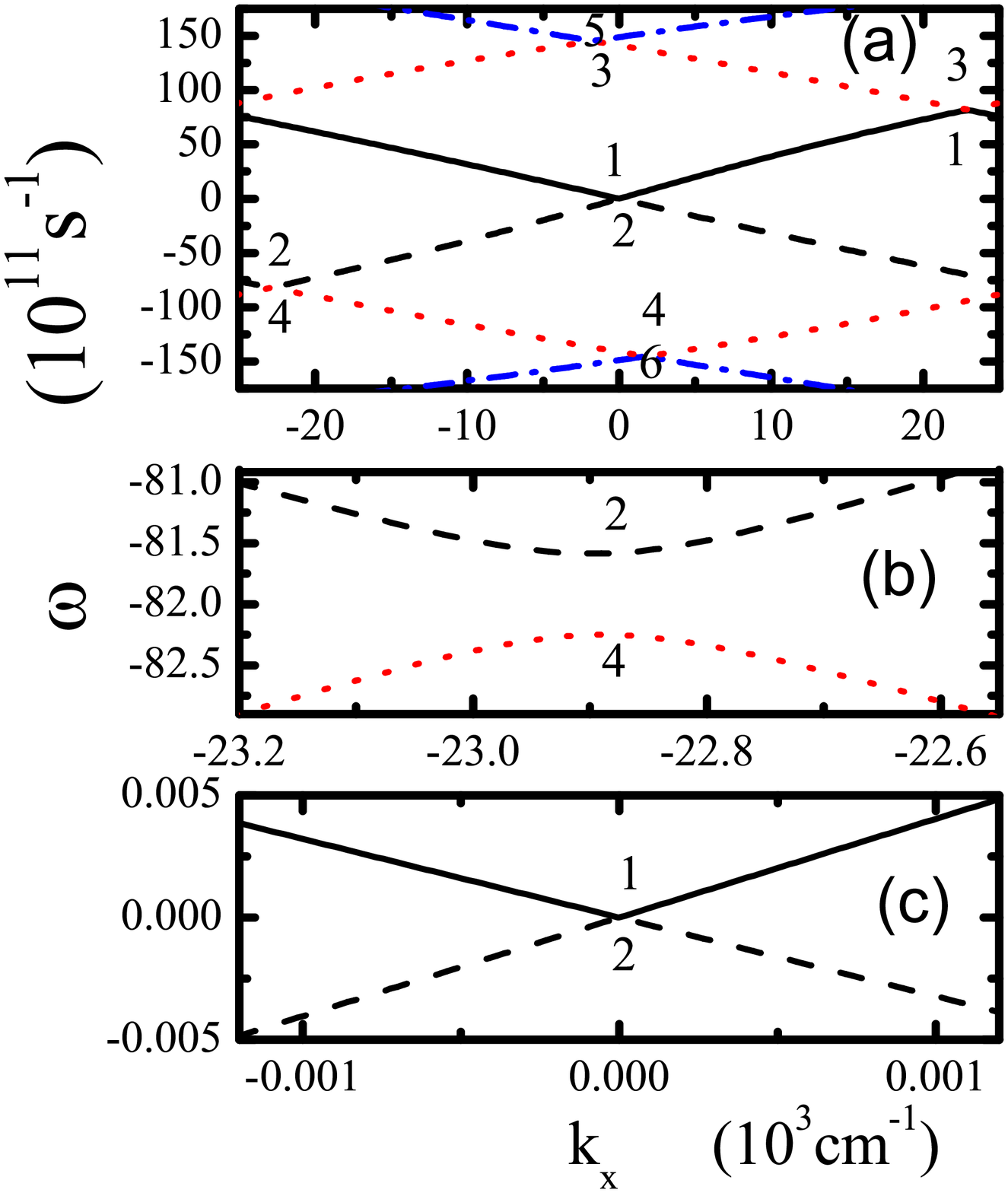}
\vspace*{-2.5cm}
\caption{(Color online) The dispersion relations $\omega(k_x, d=300$nm)
of the EMPs calculated from Eqs. (\ref{26d})-(\ref{31d})
for 
parameters of Fig. 8, except $d=300$nm.
Panel (a) presents the dispersion relations within the first Brillouin zone.
Panel (b) presents a zoom of the anticrossing for the branches 2 and 4,
at $k_{x} \approx -22900$\,cm$^{-1}$ and $\omega \approx -8.2 \times 10^{12}$\,s$^{-1}$, with the gap
$\approx 6.7 \times 10^{10}$\,s$^{-1}$.
Panel (c) presents a zoom of the branches 1 and 2
at $k_{x} \approx 0$ and $\omega \approx 0$; a finite gap is absent.
}
\end{figure}

In Fig. 4 we plot the dispersion relations $\omega(k_x,d=300$nm) of the EMPs
calculated from Eqs. (\ref{26d})-(\ref{31d}) for $d=300nm$, the rest of parameters are the same as in Fig. 2.
Fig. 4(a) shows that the branches 3 and 4 have
$\omega \approx \pm 1.6 \times 10^{11}$\,s$^{-1}$ at $k_{x} \approx 0$.
Fig. 4(b) presents a zoom of the anticrossing for the branches 2 and 4,
at $k_{x} \approx -222$\,cm$^{-1}$ and $\omega \approx -0.90 \times 10^{11}$\,s$^{-1}$, with the gap
$\approx 7.3 \times 10^{8}$\,s$^{-1}$: here the EMPs 2 and 4 have a  zero value of group velocity for pertinent $k_{x}$.
A panel for the anticrossing of the branches 1 and 3 it follows from the Fig. 4(b) by changing $k_{x}$ on $-k_{x}$
and $\omega$ on $-\omega$.
Fig. 4(c) presents a zoom of the branches 1 and 2
at $k_{x} \approx 0$ and $\omega \approx 0$; here a finite gap is absent.
The anticrossing for the branches 3 and 5
holds at $k_{x} \approx -56.7$\,cm$^{-1}$ and $\omega \approx 1.79 \times 10^{11}$\,s$^{-1}$.

In Fig. 5 we plot the dispersion relations $\omega(k_x,d\to \infty)$
of the EMPs calculated from Eqs. (\ref{26d})-(\ref{31d}) for
$2\pi/a_{0}=5000$\,cm$^{-1}$ by the curves 1-6. Fig. 5(a) presents the dispersion
relations of these EMPs within the first Brillouin zone. Notice,  Fig. 5(a) shows that the branches 3 and 4 have
$\omega \approx \pm 1.787 \times 10^{12}$\,s$^{-1}$ at $k_{x}
\approx 0$. Fig. 5(b) presents a zoom of the anticrossing for the
branches 2 and 4, at $k_{x} \approx -1860$\,cm$^{-1}$ and $\omega
\approx -1.15 \times 10^{12}$\,s$^{-1}$, with the gap $\approx 9.0
\times 10^{9}$\,s$^{-1}$: here the EMPs 2 and 4 have a  zero value
of group velocity for pertinent $k_{x}$. A panel for the
anticrossing of the branches 1 and 3 it follows from the Fig. 5(b)
by changing $k_{x}$ on $-k_{x}$ and $\omega$ on $-\omega$. Fig. 5(c)
presents a zoom of the branches 1 and 2 at $k_{x} \approx 0$ and
$\omega \approx 0$; here a finite gap is absent. Panel (d) presents
a zoom of the anticrossing for the branches 3 and 5 at $k_{x}
\approx -1112$\,cm$^{-1}$ and $\omega \approx 2.16 \times
10^{12}$\,s$^{-1}$, with the gap $\sim 5 \times 10^{6}$\,s$^{-1}$.

In Fig. 6 we plot the dispersion relations $\omega(k_x,d=3000$nm) of
the EMPs calculated from Eqs. (\ref{26d})-(\ref{31d}) for
$2\pi/a_{0}=5000$\,cm$^{-1}$. Fig. 6(a) presents the dispersion
relations of these EMPs within the first Brillouin zone by the
curves 1-6. Notice,  Fig. 6(a) shows that the branches 3 and 4 have
$\omega \approx \pm 1.783 \times 10^{12}$\,s$^{-1}$ at
$k_{x} \approx 0$. Fig. 6(b) presents a zoom of the anticrossing
for the branches 2 and 4, at $k_{x} \approx
-1920$\,cm$^{-1}$ and $\omega \approx -1.13
\times 10^{12}$\,s$^{-1}$, with the gap $\approx 8.8
\times 10^{9}$\,s$^{-1}$: here the EMPs 2 and 4 have a  zero value
of group velocity for pertinent $k_{x}$. A panel for the
anticrossing of the branches 1 and 3 it follows from the Fig. 6(b)
by changing $k_{x}$ on $-k_{x}$ and $\omega$ on $-\omega$. Fig. 6(c)
presents a zoom of the branches 1 and 2 at $k_{x} \approx 0$ and
$\omega \approx 0$; here a finite gap is absent. Fig. 6(d) presents
a zoom of the anticrossing for the branches 3 and 5 at $k_{x}
\approx -1081$\,cm$^{-1}$ and $\omega \approx
2.143 \times 10^{12}$\,s$^{-1}$, with the gap $\sim
3.7 \times 10^{6}$\,s$^{-1}$.

In Fig. 7 we plot the dispersion relations $\omega(k_x,d=300$nm) of the EMPs
calculated from Eqs. (\ref{26d})-(\ref{31d}) for $d=300nm$, the rest of parameters are the same as in Fig. 5.
Fig. 7(a) shows that the branches 3 and 4 have
$\omega \approx \pm 1.60 \times 10^{12}$\,s$^{-1}$ at $k_{x} \approx 0$.
Fig. 7(b) presents a zoom of the anticrossing for the branches 2 and 4,
at $k_{x} \approx -2220$\,cm$^{-1}$ and $\omega \approx -8.95 \times 10^{11}$\,s$^{-1}$, with the gap
$\approx 7.3 \times 10^{9}$\,s$^{-1}$: here the EMPs 2 and 4 have a  zero value of group velocity for pertinent $k_{x}$.
A panel for the anticrossing of the branches 1 and 3 it follows from the Fig. 7(b) by changing $k_{x}$ on $-k_{x}$
and $\omega$ on $-\omega$.
Fig. 7(c) presents a zoom of the branches 1 and 2
at $k_{x} \approx 0$ and $\omega \approx 0$; here a finite gap is absent.

In Fig. 8 we plot the dispersion relations $\omega(k_x,d\to \infty)$ of the EMPs
calculated from Eqs. (\ref{26d})-(\ref{31d}) for $2\pi/a_{0}=50000$\,cm$^{-1}$.
Fig. 8(a) presents the dispersion relations of these EMPs within the first Brillouin zone
by the curves 1-6. Notice,  Fig. 8(a) shows that the branches 3 and 4 have
$\omega \approx \pm 1.42 \times 10^{13}$\,s$^{-1}$ at $k_{x} \approx 0$.
Fig. 8(b) presents a zoom of the anticrossing for the branches 2 and 4,
at $k_{x} \approx -22150$\,cm$^{-1}$ and $\omega \approx -8.6 \times 10^{12}$\,s$^{-1}$, with the gap
$\approx 6.9 \times 10^{10}$\,s$^{-1}$: here the EMPs 2 and 4 have a  zero value of group velocity for pertinent $k_{x}$.
A panel for the anticrossing of the branches 1 and 3 it follows from the Fig. 5(b) by changing $k_{x}$ on $-k_{x}$
and $\omega$ on $-\omega$.
Fig. 8(c) presents a zoom of the branches 1 and 2
at $k_{x} \approx 0$ and $\omega \approx 0$; here a finite gap is absent.

In Fig. 9 we plot the dispersion relations $\omega(k_x,d=3000$nm) of the EMPs
calculated from Eqs. (\ref{26d})-(\ref{31d}) for $d=3000$nm; in addition, $2\pi/a_{0}=50000$\,cm$^{-1}$
and other parameters  coincide with those of Fig. 8.
Fig. 9(a) presents the dispersion relations of these EMPs within the first Brillouin zone
by the curves 1-6. Notice,  Fig. 9(a) shows that the branches 3 and 4 have
$\omega \approx \pm 1.42 \times 10^{13}$\,s$^{-1}$ at $k_{x} \approx 0$.
Fig. 9(b) presents a zoom of the anticrossing for the branches 2 and 4,
at $k_{x} \approx -22150$\,cm$^{-1}$ and $\omega \approx -8.6 \times 10^{12}$\,s$^{-1}$, with the gap
$\approx 6.9 \times 10^{10}$\,s$^{-1}$: notice, the parameters of this anticrossing are very close to the ones
of similar anticrossing in Fig. 8(b).
A panel for the anticrossing of the branches 1 and 3 it follows from the Fig. 9(b) by changing $k_{x}$ on $-k_{x}$
and $\omega$ on $-\omega$.
Fig. 9(c) presents a zoom of the branches 1 and 2
at $k_{x} \approx 0$ and $\omega \approx 0$; here a finite gap is absent.

In Fig. 10 we plot the dispersion relations $\omega(k_x,d=300$nm) of the EMPs
calculated from Eqs. (\ref{26d})-(\ref{31d}) for $d=300$nm; in addition, $2\pi/a_{0}=50000$\,cm$^{-1}$
and other parameters  coincide with those of Fig. 8.
Fig. 10(a) presents the dispersion relations of these EMPs within the first Brillouin zone
by pertinent curves, 1-6. Notice,  Fig. 10(a) shows that the branches 3 and 4 have
$\omega \approx \pm 1.41 \times 10^{13}$\,s$^{-1}$ at $k_{x} \approx 0$.
Fig. 10(b) presents a zoom of the anticrossing for the branches 2 and 4,
at $k_{x} \approx -22900$\,cm$^{-1}$ and $\omega \approx -8.2 \times 10^{12}$\,s$^{-1}$, with the gap
$\approx 6.7 \times 10^{10}$\,s$^{-1}$. Notice, the parameters of this anticrossing are close to the ones
in Fig. 8(b), in particular, in Fig. 10(b) the gap is only $3\%$ smaller than in Fig. 8(b).
A panel for the anticrossing of the branches 1 and 3 it follows from the Fig. 10(b) by changing $k_{x}$ on $-k_{x}$
and $\omega$ on $-\omega$.
Fig. 10(c) presents a zoom of the branches 1 and 2
at $k_{x} \approx 0$ and $\omega \approx 0$; here a finite gap is absent.
Notice, Fig. 10(a) shows that the anticrossing of the curves 3 and 5 takes place
at $k_{x} \approx 1730$\,cm$^{-1}$ and $\omega \approx 1.45 \times 10^{13}$\,s$^{-1}$.

It is natural to call in Figs. 2-10: the fundamental EMPs 1, 2 as the main fundamental EMPs, and
the fundamental EMPs 3, 4 as the first excited fundamental EMPs.
Figs. 2-10 show that for case (i), outlined in Fig.1, a strong Bragg coupling is possible due to a weak superlattice along
the edge, with the period $a_{0}$, if $L_{x}/a_{0} \gg 1$. In particular, for
frequencies in the THz range: cf. Figs. 8-10. We expect that for the frequency that corresponds to zero group velocity
of pertinent main fundamental EMP branch, or pertinent first excited fundamental EMP branch, and its vicinity the response of 
the system will have a strong resonance.

\section{ Concluding Remarks }

At the edge of a wide armchair graphene ribbon in the $\nu=2$ QHE regime and with a smooth monotonic electrostatic potential, we 
investigated the appearance of EMPs that show zero group velocity, for characteristic frequencies, 
and finite frequency gaps due to effect of a weak superlattice potential. The superlattice potential
from two original fundamental EMPs (present here without superlattice  \cite{balev2011}), due to a strong Bragg alike
coupling of them, gives correctly within as a whole first Brillouin zone two main fundamental EMPs 
(branches 1 and 2 on Figs. 2-10) and two first excited fundamental EMPs (branches 3 and 4 on Figs. 2-10).
In addition, for the wave vector within the center of the first Brillouin zone, i.e., $k_{x} \to 0$,
only the frequencies of the main fundamental EMPs, $1$ and $2$, tend to zero as the frequencies, e.g., of the first excited
fundamental  EMPs, $3$ and $4$, tend to finite values.
As at the frequency that corresponds to zero group velocity
of the main fundamental EMPs  so at one for the first excited fundamental EMPs the response of 
the system should have a strong resonance, e.g., in the THz range; see Figs. 8-10.  

Next we list and discuss the approximations used. 
Point out that in Fig. 1 (as well as in Fig. 1 of\cite{balev2011}) it is implicit that $(v_{g}(k_{x \alpha})/v_{F})^{2} \ll 1$ 
for any shown $y_{0}=\ell_{0}^{2}k_{x \alpha}$. Then extra Dirac points in the energy spectrum\cite{park2009,brey2009,barbier2010} 
due to present smooth, weak superlattices will not appear as here it follows that $\hbar v_{F} G \gg V_{s}/2$, which is opposite to 
the key condition for extra Dirac points.\cite{park2009,brey2009} 
Here for the EMPs in the $\nu=2$ QHE 
regime  dissipation is neglected, which is well justified as here the EMP damping can be
related  only with inelastic scattering processes within narrow 
temperature belts, of width $k_BT$, of each edge state 
that are much weaker than  scattering processes
due to a static disorder.\cite{balev2011} 
The latter  makes a dominant contribution to the transport scattering time in a 2DES of 
graphene \cite{novo,cast,vasko09} for $B=0$. 
Needless to say that
for a more accurate account of the EMPs studied here, dissipation must be included in the treatment.
We have neglected by nonlocal effects that usually have minor effect on
fundamental EMPs \cite{bal99}.
We emphasize that our study of the 
fundamental EMPs for the armchair termination of a graphene channel
cannot be directly extended to  zigzag termination as some important properties of the wave functions and the
energy levels are different than those of the armchair termination,  cf. \cite{cast,brey,aba,gus0,gus}. 
We relegate the study of EMPs along zigzag edges to a future work.

It is used a simple analytical model of a smooth, lateral confining potential Eq. (\ref{eq17}), however,
our main results are robust to modifications of its form and parameters if the
qualitative conditions of Fig. 1 are realized in a graphene channel in the $\nu=2$ QHE regime.
In Figs. 2-10  it is used that the exact dispersion relation $\omega(k_x,d)$
of any EMP mode can be presented in the form periodic in the reciprocal space, i.e, $\omega(k_x,d)=\omega(k_x \pm 2\pi/a_{0},d)$,
and continuous across the borders of the Brillouin zone, $\omega(\pi/a_{0}-0,d)=\omega(\pi/a_{0}+0,d)$. 
The latter, in particular, does not allow an infinite group velocity for the EMP. Point out that the dispersion 
curves of first four EMP modes, 1 - 4 , have correct periodic and continuous form
in the reciprocal space, $k_{x}$, and both qualitatively and quantitavely well describe dispersion of 
these EMP modes in graphene with the superlattice. However, an approximate dispersion curves  5 and 6 (dash-dotted) in Figs. 2-10
qualitatively correctly represent pertinent exact dependencies only nearby the anticrossings of 5 with 3 and of 6 with 4.
So the curves 5 and 6 are shown only within a small part of the first Brillouin zone in Figs. 2(a)-10(a). 
In addition, as the second order contributions over the periodic
potential are neglected (as well as an additional contributions in Eqs. (\ref{19d}) with the $\ell=\pm 2$) the 
Figs. 2(d), 3(d), 5(d), 6(d) give only rough approximation for this anticrossing and, in particular, for its gap.

\begin{acknowledgments}
O. G. B. acknowledges support by Brazilian FAPEAM
(Funda\c{c}\~{a}o de Amparo \`{a} Pesquisa do Estado do Amazonas)
Grant and by the Brazilian Council for Research (CNPq)
APV Grant No. 452849/2009-8. A. C. A. Ramos thanks the FUNCAP (Funda\c{c}\~{a}o Cearense de Apoio
ao Desenvolvimento Cient\'{i}fico e Tecnol\'{o}gico) for financial
support. 
\end{acknowledgments}
\appendix

%


%

\begin{thebibliography}{10}
%
%
\bibitem{novo} K. S.~Novoselov, A. K.~Geim, S. V.~Morozov, D.~Jiang, Y.~Zhang, S. V.~Dubonos, I. V.~Grigorieva, and A. A.~Firsov, Science {\bf 306}, 666 (2004);
K. S.~Novoselov,   {\it Proc. Natl. Acad. Sci. USA} {\bf 102}, 10451 (2005);
A. K.~Geim and K. S.~Novoselov, Nature Materials, {\bf 6}, 183 (2007).


\bibitem{cast} A. H.~Castro Neto, F.~Guinea, N. M. R.~Peres, K. S.~Novoselov, and A. K.~Geim, Rev.~Mod.~Phys.~{\bf 81}, 109 (2009).

\bibitem{wallace} P. R. Wallace,   Phys. Rev. {\bf 71}, 622 (1947).

\bibitem{klein} O. Klein, Z. Phys. {\bf 53}, 157 (1929).

\bibitem{kat} M. I. Katsnelson, K. S. Novoselov, A. K. Geim,
Nature Phys. {\bf 2}, 620 (2006); J. Milton Pereira Jr.,  P. Vasilopoulos, and F.
M. Peeters, Appl. Phys. Lett. {\bf 90}, 132122, (2007).

\bibitem{brey} L. Brey and H. A. Fertig,   Phys. Rev. B {\bf 73}, 195408 (2006);
N. M. R. Peres, F. Guinea, A. H. Castro Neto, {\it ibid} {\bf 73}, 125411 (2006).

\bibitem{aba} D. A. Abanin, P. A. Lee, and L. S. Levitov, Phys. Rev. Lett. {\bf 96}, 176803 (2006);
Solid State Commun., {\bf 143}, 77 (2007).

\bibitem{gus0} V. P. Gusynin, V. A. Miransky,  S. G. Sharapov, and I. A. Shovkovy,  Phys. Rev. B {\bf 77}, 205409 (2008).

\bibitem{zit1} N. M. R. Peres,  A. H. Castro Neto, F. Guinea, Phys. Rev. B {\bf 73}, 241403 (2006).


\bibitem{zit2} H.-Y. Chen, V. Apalkov, and T. Chakraborty, Phys. Rev. Lett.  {\bf 98}, 186803 (2007);
 A. V. Shytov, M. S. Rudner, and L. S. Levitov, {\it ibid} {\bf 101}, 156804 (2008).

\bibitem{park2009} C.-H. Park, Y.-W. Son, Li Yang, M. L. Cohen, and S. G. Louie,   Phys. Rev. Lett. {\bf 103}, 046808 (2009).

\bibitem{brey2009} L. Brey and H. A. Fertig,   Phys. Rev. Lett. {\bf 103}, 046809 (2009).

\bibitem{barbier2010} M. Barbier,  P.~Vasilopoulos, and F. M. Peeters, Phys. Rev. B {\bf 81}, 075438 (2010);
M. Barbier, F. M. Peeters,  P.~Vasilopoulos, and J.M. Pereira, {\it ibid} {\bf 77}, 115446 (2008).

\bibitem{gus} V. P. Gusynin, V. A. Miransky,  S. G. Sharapov, I. A. Shovkovy, and C. M. Wyenberg,  Phys. Rev. B {\bf 79}, 115431 (2009).

\bibitem{milt} J.M. Pereira, F. M. Peeters, and P.~Vasilopoulos, Phys. Rev. B {\bf 75}, 125433 (2007).


\bibitem{balev2011} O.G. Balev, P. Vasilopoulos, and H. O. Frota, Phys. Rev. B {\bf 84}, 245406 (2011).

\bibitem{volkov91} V.A. Volkov and S.A. Mikhailov, ``Electrodynamics of Two-Dimensional Electron Systems in
High Magnetic Fields,'' in Landau Level Spectroscopy, Modern Problems in Condensed Matter Sciences,
Ed. by G. Landwehr and E. I. Rashba (North-Holland, Amsterdam,
1991), vol. 27.2, ch.15, p. 855-907; V.A. Volkov and S.A. Mikhailov, Zh. Eksp. Teor. Fiz. {\bf 94}, 217 (1988)
[Sov. Phys. JETP {\bf 67}, 1639 (1988)].

\bibitem{aleiner94} I. L. Aleiner and L. I. Glazman, Phys. Rev. Lett. {\bf 72}, 2935 (1994).


\bibitem{wen91} B. I. Halperin, Phys. Rev. B {\bf 25}, 2185 (1982);
X. G. Wen, {\it ibid} {\bf 43}, 11025 (1991);
M. Stone, Ann. Phys. (N.Y.) {\bf 207}, 38 (1991).

\bibitem{stone92} M. Stone, H. W. Wyld, and R. L. Schult, Phys. Rev. B {\bf 45}, 14156 (1992);
U. Zulicke and A. H. MacDonald, {\it ibid} {\bf 54}, 16813 (1996);
S. Giovanazzi,L. Pitaevskii, and S. Stringari, Phys. Rev. Lett. {\bf 72}, 3230 (1994).

\bibitem{bal} O. G. Balev and P.~Vasilopoulos, Phys. Rev. Lett. {\bf 81}, 1481 (1998); O.G. Balev, P. Vasilopoulos, and
Nelson Studart, J. Phys.: Condens. Matter {\bf 11}, 5143 (1999);
O. G. Balev and P.~Vasilopoulos, Phys. Rev. B {\bf 56}, 13252 (1997).

\bibitem{bal2000} O. G. Balev and Nelson Studart, Phys. Rev. B {\bf 61}, 2703 (2000);
Sanderson Silva and O. G. Balev, J. Appl. Phys. {\bf 107}, 104310 (2010).


\bibitem{bal99} O. G. Balev and P.~Vasilopoulos, Phys. Rev. B {\bf 59}, 2807 (1999).



\bibitem{ashoori92} R. C. Ashoori, H. L. Stormer, L. N. Pfeiffer, K. W. Baldwin, and K. West,
Phys. Rev. B {\bf 45}, 3894 (1992).

\bibitem{ernst96} G. Ernst, R. J. Haug, J. Kuhl, K. von Klitzing, and K. Eberl, Phys. Rev. Lett. {\bf 77}, 4245 (1996).

\bibitem{kukushkin09} M. N. Khannanov, A. A. Fortunatov, and I. V. Kukushkin,
Pis'ma Zh. Eksp. Teor. Fiz. {\bf 90}, 740 (2009) [JETP Lett.{\bf 90}, 667 (2009)].

\bibitem{bal2000b} O. G. Balev,  Nelson Studart, and P. Vasilopoulos, Phys. Rev. B {\bf 62}, 15834 (2000).

\bibitem{landaulif} L.D. Landau and E.M. Lifshitz, {\it Quantum Mechanics} (Pergamon Press, New York, 1975).

\bibitem{bee} C. W. J. Beenakker and H. van Houten, in {\it Quantum Transport in
Semiconductor Nanostructures}, Solid State Physis Vol. 44
edited by H. Ehrenreich and D. Turnbull (Academic, San Diego, 1991).

\bibitem{thouless93} D. J. Thouless, Phys. Rev. Lett. {\bf 71}, 1879 (1993).

\bibitem{bal96} O. G. Balev and P.~Vasilopoulos, Phys. Rev. B {\bf 54}, 4863 (1996).


\bibitem{zhe} Y. Zheng and T. Ando, Phys. Rev. B {\bf 65}, 245420 (2002); V. P. Gusynin and S. G. Sharapov,
Phys. Rev. Lett. {\bf 95}, 146801 (2005).

\bibitem{vasko09} O. G. Balev, F. T. Vasko, and V. Ryzhii, Phys. Rev. B {\bf 79}, 165432 (2009).

\end{thebibliography}
\end{document}